\pdfoutput=1

\documentclass[
twocolumn,times,tighten,twocolappendix,
]{aastex631}

\usepackage{graphicx,graphics,amsmath}
\usepackage{natbib}
\usepackage{bm,url}
\usepackage{color}
\usepackage{array} 
\usepackage{multirow}
\usepackage{hyperref}
\usepackage{float}
\usepackage{csvsimple}
\usepackage{enumitem}  
\usepackage{booktabs} 

\graphicspath{{./figs/}{./png/}}

\def\bl{Babcock--Leighton}
\newcommand{\Fig}[1]{Figure~\ref{#1}}
\newcommand{\Figs}[2]{Figures~\ref{#1} and \ref{#2}}

\newcommand{\Eq}[1]{Equation~(\ref{#1})}
\newcommand{\Sec}[1]{Section~\ref{#1}}
\newcommand{\Tab}[1]{Table~\ref{#1}}

\newcommand{\mps}{m~s$^{-1}$}

\newcommand{\II}{{\sc \romannumeral 2} }

\emergencystretch=\maxdimen
\shorttitle{STELLAR CYCLE VARIABILITY}
\shortauthors{GARG et al.}

\begin{document}

\title{Stellar cycle variability in Mount Wilson stars and dynamo models: Rotation rate and dynamo number dependency}

\author[0000-0002-2309-9731]{Suyog Garg}
\affil{Department of Physics, The University of Tokyo, Bunkyo-ku, Tokyo 113-0033, Japan}
\affil{Research Center for Early Universe, The University of Tokyo, Bunkyo-ku, Tokyo 113-0033, Japan}

\correspondingauthor{Bidya Binay Karak}
\email{karak.phy@iitbhu.ac.in}

\author[0009-0002-9274-6872]{Rohan B. Mandrai}

\author[0000-0002-8883-3562]{Bidya Binay Karak} 
\affil{Department of Physics, Indian Institute of Technology (Banaras Hindu University), Varanasi 221005, India}

\begin{abstract}
Similar to the solar cycle, the magnetic cycles of other solar-type stars are also variable. 
How the variability of the stellar cycle changes with the rotation rate or the dynamo number is a valuable information for understanding the stellar dynamo process. We examine the variability in the stellar magnetic cycles by studying 81 stars from the data of the Mount Wilson Observatory, which started observations in 1966. For 28 stars, we have time series data available till 2003, while for others, the data are limited till 1995. We specifically explore how the variability changes with respect to three rotation-related parameters. We find a modest positive correlation between the variability and the stellar rotation period. In addition, we find suggestive negative correlations between the variability and the inverse squared Rossby number ($Ro^{-2}$), and the ratio of the mean cycle duration and rotation period ($\log \, (\langle P_{\rm cyc} \rangle / P_{\rm rot})^2$). Variability computed from the magnetic field of stellar dynamo models also show similar trends. Finally, inspired by previous studies, we examine dynamo number scaling in our model data and find that $Ro^{-0.6}$ (instead of $Ro^{-2}$ as suggested in the linear $\alpha \Omega$ dynamo theory) and $(\langle P_{\rm cyc} \rangle /P_{\rm rot})^{0.6}$ (instead of $\log \, (\langle P_{\rm cyc} \rangle / P_{\rm rot})^2$ as predicted in previous observations) are a good measure of the dynamo number. In conclusion, our results demonstrate that the stellar magnetic cycle variability decreases with the increase of the rotation rate or the dynamo number.
\end{abstract}

\keywords{Sun: activity -- (Sun:) sunspots -- Sun: magnetic fields -- Sun: interior  -- magnetohydrodynamics (MHD) -- dynamo}


\section{Introduction}
\label{sec:intro}

Solar activity shows a spectrum of variability besides the 11-year periodicity in the magnetic field. Variability ranges from changes in activity at a small scale, such as quasi-biennial oscillation and Gnevyshev/double peaks, to the long-term modulation and grand minima and maxima \citep{LB89, KMB18, Biswas23}.
The magnetic field of other solar-type stars also shows cyclic behavior that depends on various properties of the stars \citep{Baliu95}. The magnetic field (or its proxy, such as chromospheric and coronal emissions) increases with decreasing rotation period, and the activity tends to saturate below a rotation period of about 10 days \citep{Skumanich72, R84, Noyes84a, wright11}. 
The observations of chromospheric emission during 1966--2003 at Mount Wilson Observatory (MWO) reveal further properties of stellar cycles, (i) the rapidly rotating young stars produce high magnetic activity levels and irregular cycles, (ii) stars of intermediate age display moderate levels of activity and rotation rates and occasional smooth cycles, and (iii) stars as old as the Sun and older show slow rotation, weak activity levels, smooth cycles and occasional Maunder minimum-like phases \citep{Baliu95, Egeland:2017:thesis}. These general trends are also supported by other related observations \citep{NM07, Ol16, BoroSaikia18, Garg19}.    Recently, \cite{Shah18, Baum22, luhn22, Baum22} confirmed the earlier prediction of the Maunder minimum-like phase in slowly rotating stars with new data.

In addition to the properties of stellar cycles differing across the age and rotation rate of different stellar populations, the cycle properties also vary within the time series data of a single star.  For the Sun, we have seen that the solar cycle amplitude has a significant variation from one cycle to another, making the cycle prediction more challenging \citep{petrovay20, Kumar22, Kar23}.  Unfortunately, for other stars, we have observation data for only a limited duration.  Despite this, even from these limited observations as presented in Fig 1.\  of \cite{Baliu95}, it is observed that the cycle amplitude in a star has a significant variation, and the amount of this variation is different for different stars. \cite{SB02} studied how the normalized peak-to-peak variation in the cycle activity changes with  the ratio of cycle duration and rotational frequencies for some MWO stars.
How the amount of this cycle variation changes with the stellar rotation rate is one of the motivations of the present work. 

To understand the dynamo operation in various stars, it is essential to identify this dependency of stellar cycle variability with the rotation rate of stars from observations. 
The dynamo process is possible when the dynamo number ($D$) is above a particular value, called the critical dynamo number $D_c$ \citep{kr80, cha20}.
In other words, for $D$ above this $D_c$, the magnetic field is maintained or even oscillates through the dynamo process.  
Near the dynamo transition, the influence of the nonlinearity is weak, and the growth rate of the magnetic field is small.  
Hence, due to fluctuations, if $D$ increases, then the change in the magnetic field will not be much as the dynamo will take a long time to grow the magnetic field to a high value and by that time, a new fluctuation can alter the dynamo number.  On the other hand, if $D$ is much larger than $D_c$ (highly supercritical regime), the dynamo growth rate is large, and the dynamo quickly amplifies the magnetic field to a large value if $D$ is increased to a large value due to fluctuations. Hence the stellar cycle variability should increase with the increase of dynamo number.  However, we shall find an opposite trend if there is a strong nonlinear quenching effect in the dynamo, which reduces the dynamo growth in the highly supercritical regime (see, e.g., Fig. 16 of \citet{Kar23} and compare it with Figs. 1b-c of \citet{kumar21b}). Finally, a large and complex behaviour can be obtained when the dynamo solution changes to highly irregular or chaotic at large $D$, as demonstrated with a specific type of nonlinearity by \citet{CSZ05} and \citet{TC23}.   Hence, it is crucial to identify the trend of the stellar cycle variability with the rotation rate and dynamo number from observations and compare it with the model.

The layout of the paper is as follows. In \Sec{sec:obsdata}, we introduce the data and the analyses methods for this study. After that we present our dynamo models in \Sec{sec:moddata}. Then in \Sec{sec:result}, we shall discuss our results from the observational data, and compare them with those from our dynamo models (\Sec{sec:res-mod}). We conclude in \Sec{sec:conclu} with some final comments.



\section{Observation Data and Analysis}
\label{sec:obsdata}

\begin{figure*}[ht]
    \centering
    \includegraphics[width=1.\textwidth]{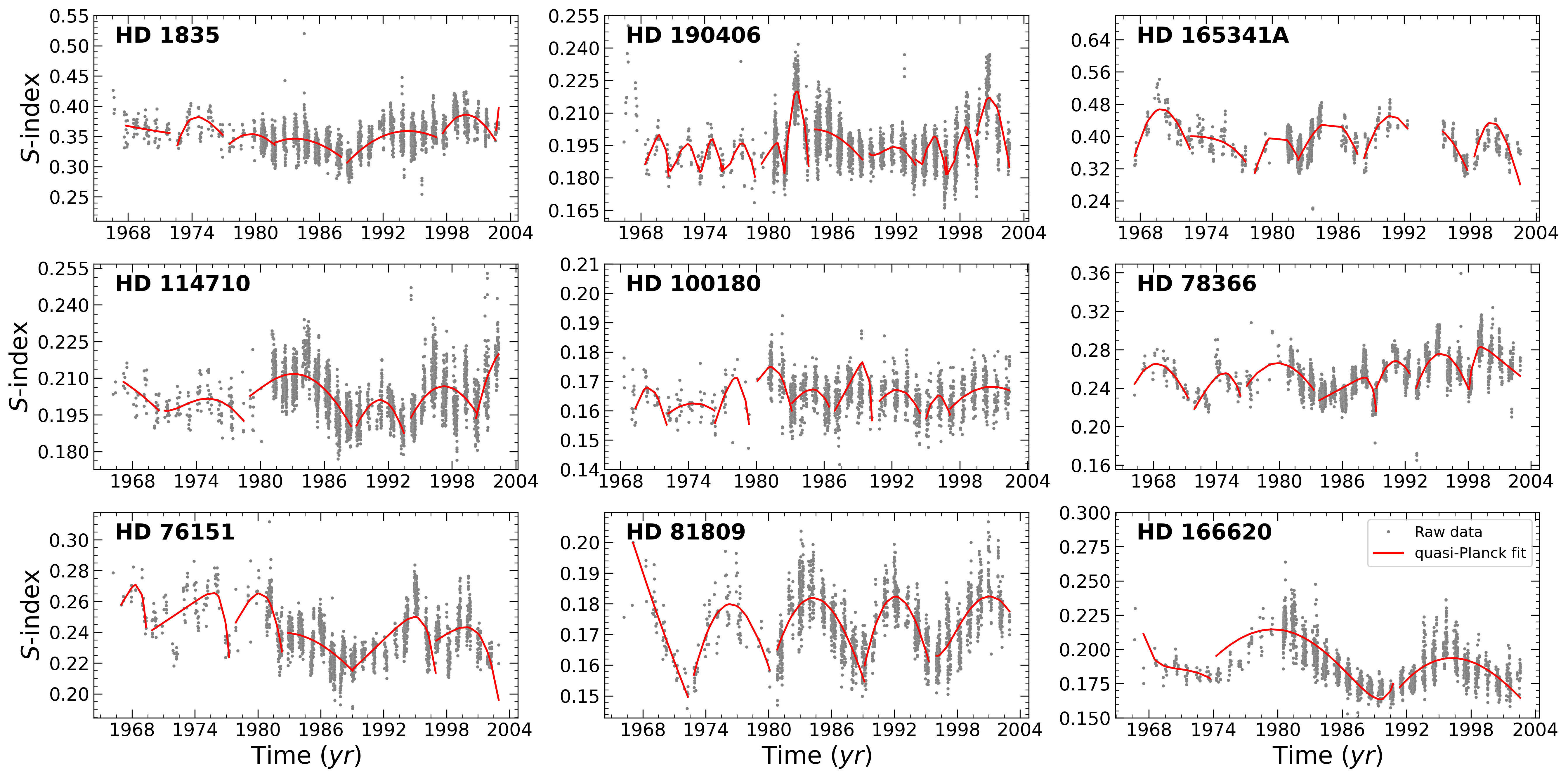}
    \caption{Time series plots of MWO raw $S$-index data and the respective quasi-Planck fittings for the seven additional GK25 stars considered in this study. Two stars, HD~81809 and HD~166620, that have relatively well defined cycles, and which were studied by \cite{Garg19}, are also shown for cycle comparison. Properties of the stars are tabulated in \Tab{tab:data}, while the fit parameters are given in \Tab{tab:fitParams}.}
    \label{fig:timeseries}
\end{figure*}


\subsection{MWO $S$-index Data and Cycle Fittings}
\label{sec:mwodata}

For 37 years between 1966$-$2003, the Mount Wilson Observatory (MWO) measured the stellar chromospheric activity for more than 2290 stars using its H-K photospectrometer. The data from these observations are available in the form of the Ca \II H and K line emission cores, from which the MWO $S$-index value is defined. The $S$-index is proportional to the ratio of measured flux within the H and K bandpasses of the MWO photometer to that of the two continuum windows equidistant to the red and violet of the H and K lines.

\cite{Garg19} (hereafter GK19) selected 21 stars from the MWO dataset which showed solar-like cycle periodicity and analyzed their $S$-index time series data. The selection was also motivated by the list of stars studied by \cite{BV07} and \cite{Sch13}, because they were identifying the Waldmeier effect \citep{wald, KC11} which required stellar cycles with well identified peaks and rising phase.
In the present work, we want to investigate variabilities in stellar activity encapsulated in the MWO chromospheric activity data.
We take the GK19 stars as our main stars. Now, since we are not constrained to only consider fully complete cycles, we are also able to include additional partial cycles for most of these stars, except HD~149661, HD~161239, and HD~219834B. These partial cycles are usually located at the start or the end of the time series data.

Moreover, when studying the variability in the long-term data, we are not interested in individual cycle peaks. That is, we do not necessarily need stars to have full well-defined (easily identifiable peaks) cycles in their time series data. Therefore, we can also include some additional stars from the MWO dataset on top of the GK19 stars, namely: HD~1835, HD~76151, HD~78366, HD~100180, HD~114710, HD~165341A, HD~190406.
These stars do not necessarily have well-defined cycle peaks, however, they have relatively high-quality data available till 2003. These seven stars, together with the 21 GK19 stars, make up our GK25 set.

 In the Appendix, we provide the list of properties for all the stars considered in our analysis, along with the number of full and partial cycles for each of them.
To compute the cycle period of any star, we follow the same procedure as GK19. First, we take an yearly bin of the raw $S$-index data and make a smoothed time series. We then find the times at which the rate of change in this yearly-binned activity changes polarity, e.g., from a negative rate (declining phase) to positive (rising phase). Using these $n$ minima, we can then divide the data into $(n-1)$ cycles, with one stellar cycle comprising the data between two consecutive minima.
Computing the cycle periods in this way also enables us to define the cycle start and end times effectively. This, in turn, aids in fitting  smoothing functions to the data.
The mean cycle duration, $\langle P_{\rm cyc} \rangle$, is then calculated as the average value of the period of all such usable cycles, while partial and ill-defined cycles are ignored. 
Here, we are not considering the secondary cycles which are also seen in many stars (see \citet{SB02}).

Following GK19, we also fit the following  quasi-Planck function \citep{hathaway1994} to the yearly binned MWO $S$-index data:
\begin{equation}
f(t) = \dfrac{a\, (t-t_0)^3}{ \exp\left[ \dfrac{(t-t_0)^2}{b^2} \right] - c }
\label{eq:planck}
\end{equation} where, $t$ is the time in years, $t_0$ is the cycle start time, and the parameters $a$, $b$ and $c$ can be thought of as relating to the cycle amplitude, rise time and asymmetry respectively. 
For simplicity we only consider the quasi-Planck fitting in the present study, and not the skewed-Gaussian fitting that GK19 also considered.
They also give the cycle fit parameters for all the GK19 stars, including those for the partial cycles they did not later consider, which we directly use.
However, for the seven additional stars, we perform the fitting anew. 
The fit parameters and cycle properties for these are given in the Appendix.
\Fig{fig:timeseries} illustrates the time series data for these additional stars and two representative GK19 stars. Partial cycles and the quasi-Planck fittings are also shown.

Now, the range of rotation period for the GK19 stars is limited to above $P_{\rm rot}>10$ days. 
The seven new stars additionally considered here, do have shorter periods of rotation, but are limited to just a small number.
Thus, to include more number of rapidly rotating stars, that have rotation period less than 10 days, we look up the set of 104 MWO stars studied by \cite{Baliu95} (B95).
From this set, we select those stars that have rotation periods and Rossby numbers available in Table 5.3 of \cite{Egeland:2017:thesis}, which lists properties of many MWO stars together with their variability classes.
This gives us 53 other B95 stars (apart from the GK25 set) to consider. The properties for these stars are also detailed in the Appendix.
These stars are characterized by a range of variability behaviour, from flat non-cyclic to excellent cyclic stellar activity. However, proper cyclic behaviour is not easily discernible for all stars that were not already studied by GK19. Thus, we do not perform the quasi-Planck fit for any of these stars and only consider their raw data for the analysis.
Further, the GK25 stars form a subset of B95 set of stars, but have longer data available (till 2003) compared to the rest of the stars (which only have data till 1995).



Lastly, we note that the MWO $S$-index is sensitive to the integrated emission within the H and K bandpass window, which depends on the spectral type of the star. It is also sensitive to the photospheric radiation that gets included in the chromospheric emission measurement by the MWO H-K photometer. 
\cite{Noyes84a} outline methods to correct for both of these effects, by parametrizing the MWO $S$-index into the chromospheric emission ratio $R^{\prime}_{\rm HK}$ using a B--V color dependent conversion factor $C_{\rm ef}$ and the photospheric emission component $R_{\rm photo}$. We also follow the same procedure and convert our $S$-index data to $R^{\prime}_{\rm HK}$, while performing the correction in $C_{\rm ef}$ only for stars with $b_v<0.63$: HD~78366, HD~100180, HD~114710 and HD~190406; and using the cubic equation in $b_v$ for the $R_{\rm photo}$ definition. The $b_v$ column in \Tab{tab:data} tabulates the B--V color amplitude for all stars considered in this study.

\subsection{Variability}
\label{sec:var}

We define the ``variability" in the parametrized stellar activity data as the quantity:
\begin{equation}
    \rm{var(R)} = \dfrac{\Delta R^{\prime}_{\rm HK}}{\langle R^{\prime}_{\rm HK} \rangle}
\label{eq:var-rhk}
\end{equation}%
where, the numerator denotes the standard deviation in $R^{\prime}_{\rm HK}$, while the denominator is the mean value.

Now, we can have two approaches when calculating the variability in the MWO stellar data: use the cycle peak values and find how they vary over time, or use the full raw $S$-index data.
For a majority of the MWO stars, figuring out the cycle maxima is tricky, since either the behavior is non-cyclic to begin with or clearly defined cycles are non-existent. Thus, we have well-defined cycle peak values for only the GK19 stars.
In spite of this, \cite{SB02} did analyze changes in normalized peak-to-peak cycle variability with respect to the ratio of cycle duration and rotation period and found the existence of two parallel branches for stars in their studied dataset
(note that they also included some stars present in GK25 set for their analysis).
However, we still feel that the number of stars with well-defined cycles are too low in number to enable any significant correlation evaluation. 
Therefore, we shall use the full raw $S$-index data to compute the variability.

The MWO raw $S$-index data available for use starts in 1966 and goes on till 2003 for the GK25 (or till 1995 in B95) set of stars. The $S$-index data is not averaged out in any manner and is instead directly converted to $R^{\prime}_{\rm HK}$ activity level. When considering the quasi-Planck fit, the data from the individual cycle fits are first combined together, and then the variability is calculated on these concatenated data, a process akin to smoothing out of the raw observation data.
The $\rm{var(R)}$ and $\Delta R^{\prime}_{\rm HK}$ values for all the stars used in the analysis are also provided in the Appendix.
In \Sec{sec:result}, we shall discuss the correlations observed between the stellar variability and different cycle properties.


\section{Stellar Dynamo Model and Cycles}
\label{sec:moddata}
To compare the stellar cycle variability obtained in the MWO data with the theoretical stellar cycles, we shall utilize the dynamo models of \citet{VKK23}. While the details of these dynamo models are presented in previous publications \citep{VKK23, Hazra19, KKC14}, here we mention some salient features.  These models solve the equations for the kinematic dynamo under the axisymmetric approximation. 
The large-scale flows (differential rotation and meridional circulation) are obtained from the hydrodynamic model of \citet{KO11b} and \citet{KO12b}. In this model, the structure of the star is specified following the stellar evolution code of \citet{Paxton} for $1M_\odot$ star as a function of age. 
Then, from age, the rotation rate is computed using the Gyrochronology relation of \citet{B07}. Essentially, the model provides differential rotation and meridional circulation in the Sun at different ages or the corresponding rotation periods of 1, 2, 3, 4, 5, 7, 10, 15, 20, 25.38 (solar value), and 30 days.
Finally, using these flows, the dynamo models are set up for these stars rotating at different periods.
We perform dynamo simulations for three cases which are labelled as Models I--III. In model I, the diffusivity is low ($5 \times 10^{10}$~cm$^2$~s$^{-1}$ in CZ and $2 \times 10^{12}$~cm$^2$~s$^{-1}$ in top 5\% of the star) while Model II is the same as Model I but for the full sphere. Finally, Model III is the same as Model II but in this case, the diffusivity is high ($3 \times 10^{11}$~cm$^2$~s$^{-1}$ in CZ and $3 \times 10^{12}$~cm$^2$~s$^{-1}$ in top 5\% of the star) and a downward magnetic pumping (of speed 24~\mps) is included, to make it consistent with observations and surface flux transport model \citep{KC16, KM17}.

\begin{table}[ht]
\begin{center}
\caption{Spearman rank correlation coefficients between ${\rm var(R)}$ and rotation related parameters for different data cases.}
\label{tab:corr}
\begin{tabular}{ll*{3}{c}}
\toprule
   Stars
    & $N$
    &  $P_{\rm rot}$
    &  $(Ro)^{-2}$
    &  $\log \, (\langle P_{\rm cyc} \rangle / P_{\rm rot})^2$ \\
\midrule\midrule
\cmidrule{2-5}
\textbf{GK19} 
    & 21 & 0.44 & $\mathbf{-0.53}$ & $-0.20$ \\
GK19p
    & & 0.50 & $-0.45$ & $-0.32$ \\
\textbf{GK19$+$}
    & 22 & 0.47 & $-0.37$ & $-0.28$ \\
GK19p$+$
    & & \textbf{0.53} & $-0.29$ & $-0.38$ \\
\midrule
\textbf{GK25} 
    & 28 & 0.31 & $-0.40$ & $-0.36$ \\
GK25p
    & & \textbf{0.53} & $-0.33$ & $-0.36$ \\
\textbf{GK25$+$} 
    & 29 & 0.31 & $-0.28$ & $-0.39$ \\
GK25p$+$
    & & \textbf{0.52} & $-0.22$ & $-0.38$ \\
\midrule
\textbf{B95}
    & 81 & \textbf{0.51} & $\mathbf{-0.30}$ & $\mathbf{-0.66}$ \\
B95$+$ 
    & 82 & \textbf{0.50} & $\mathbf{-0.28}$ & $\mathbf{-0.65}$ \\
\bottomrule
\end{tabular}
\end{center}
\vspace{-0.25cm}
\tablecomments{GK19 stars are from \cite{Garg19}. GK25 set includes 7 additional stars on top of GK19. B95 refers to the list of stars studied by \cite{Baliu95} and includes all GK25 stars plus some more fast rotators with irregular cycles. The $+$ means that the solar data is additionally included in the set, while $p$ denotes use of quasi-Planck fit data. $N$ is the base number of stars in each set, however, only stars with data available in \Tab{tab:data} are used in different correlation evaluations. 
The values in bold indicate significant correlations having $p\le 0.01$.
}
\end{table}

The stellar chromospheric emission as measured here by $S$ index is caused by the surface magnetic field of the star, which is largely radial. As the emission is expected to be caused by the magnetic reconnection of one flux system with another, and we naively expect the emission to depend on the flux of the unsigned surface magnetic field \citep[see for example][]{Viodoto14}. Therefore, to compute the magnetic field variability from the stellar dynamo models, we first compute the radial magnetic field $B_r(R_s, \theta, t)$ on the solar surface at every month from each simulation of 11,000 years long.
Next, we compute the total flux of the unsigned radial magnetic field ($|B_r(R_s, \theta, t)|$) on the entire surface of the star, which we call $\Phi_r (t)$. Then, we compute the variability in the same way as done for the observed data (\Eq{eq:var-rhk}), namely,
\begin{equation}
    {\rm var(\Phi)} = \dfrac{\Delta \Phi_r}{\langle \Phi_r \rangle}.
    \label{eq:modvariability}
\end{equation}%
We note that $\Delta \Phi_r$ and ${\langle \Phi_r \rangle}$ are computed from the whole time series of 11,000 year long simulation.
Finally we compute the periods of the dominant cycles in the same way as done in \citet{VKK23}. We recall that our dynamo models do not produce secondary (short) cycles which are frequently generated in the some global simulations of stellar dynamo \citep{Kap16}.


\section{Results and Discussion}
\label{sec:result}

\subsection{Variability in observed data}
\label{sec:res-obs}

\begin{figure}[ht]
    \centering
    \includegraphics[width=\linewidth]{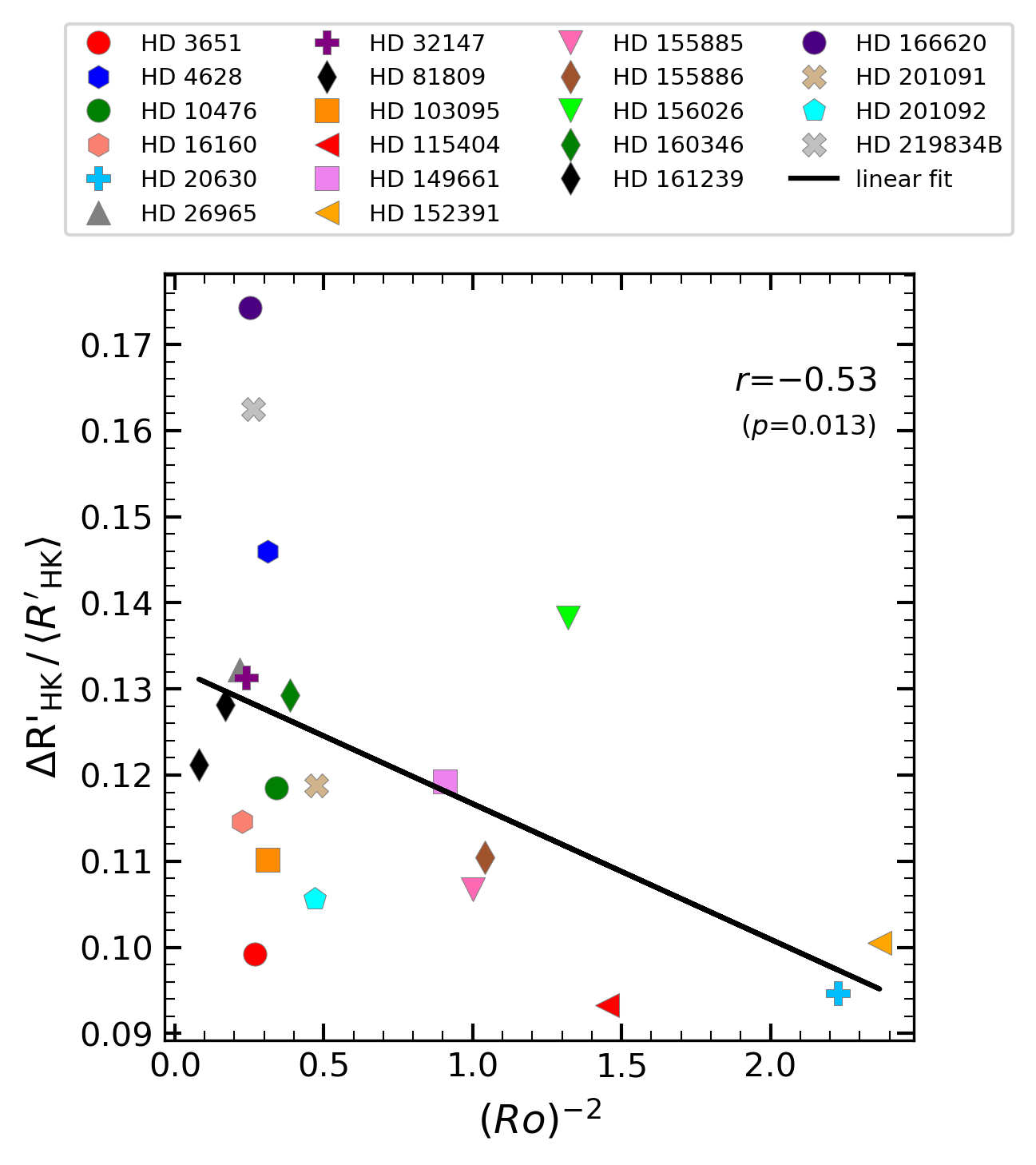}
    \caption{Scatter plot between the parametrized variability (\Eq{eq:var-rhk}) and the inverse squared Rossby number for the GK19 set of stars. The Spearman rank correlation $r$ is also provided.}
    \label{fig:gk19rossby}
\end{figure}

\begin{figure}[ht]
    \centering
    \includegraphics[width=\linewidth]{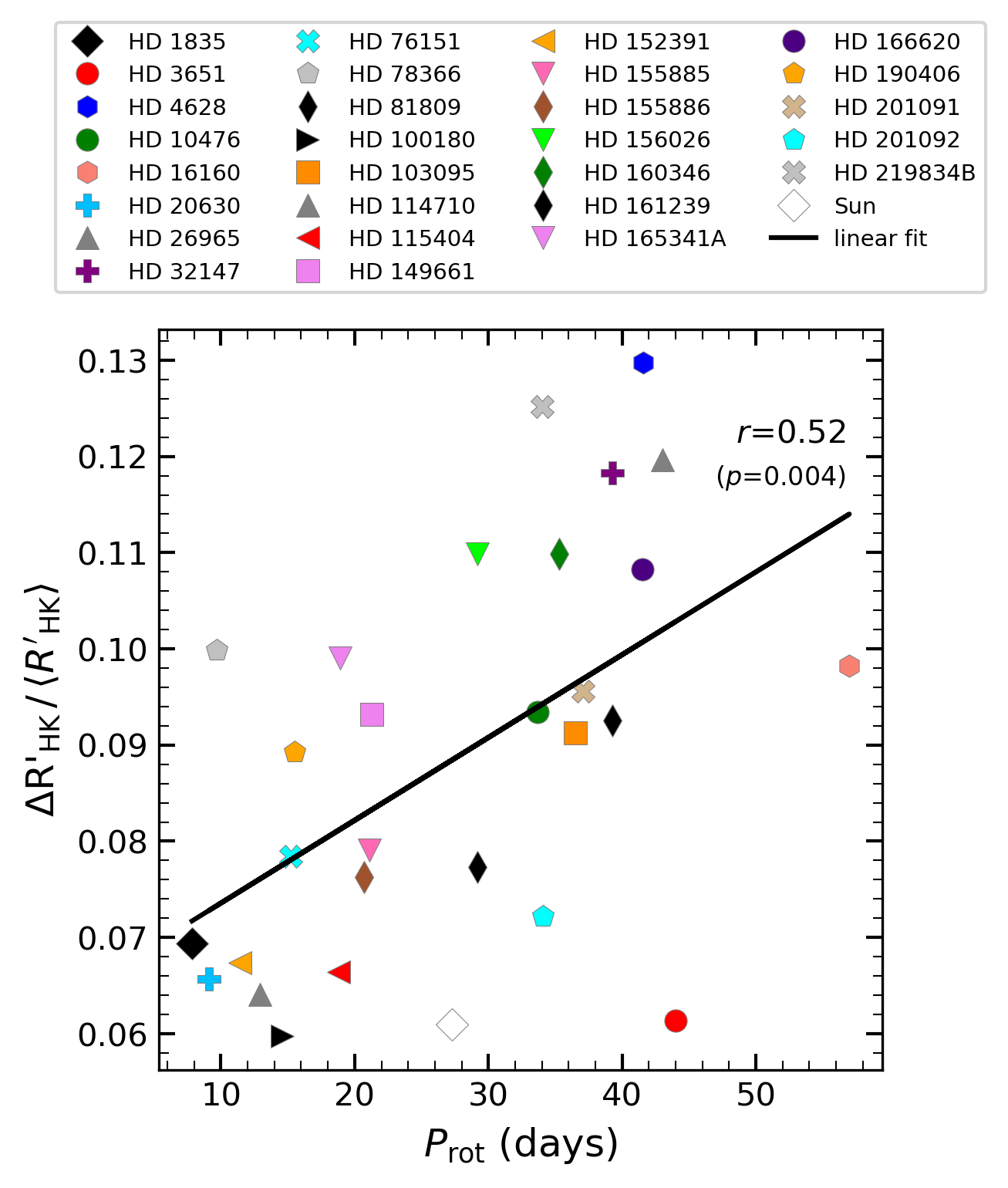}
    \caption{Same as \Fig{fig:gk19rossby} but for evaluating the variability and rotation period correlation of GK25$+$ set of stars using the quasi-Planck fit data.}
    \label{fig:gk25p+prot}
\end{figure}

The stellar magnetic activity is measured via the MWO $S$-index and its parametrized form $R^{\prime}_{\rm HK}$. The variability in the activity can be quantified in terms of the quantity var($R$), as defined in \Sec{sec:obsdata}.
We now want to investigate how this stellar variability changes with the rotation rate of the stars.
It is expected that with increasing rotation rate, the dynamo efficiency measured by the dynamo number becomes stronger due to the strengthening of the helical nature of convection ($\alpha$ effect) and the weakening of turbulent diffusion. However, there are other factors, such as the amount of shear and the depth of CZ, that determine the dynamo number.  \cite{DL77} showed that the inverse squared Rossby number $(Ro)^{-2}$ is a measure of the dynamo number in the linear $\alpha \Omega$ dynamo theory (where $Ro$ is a ratio of the rotation period and the convective turnover time as given by \citet{Gil80}); also see \citet{Noyes84a}.  On the other hand, \cite{SBZ93} showed that the ratio of stellar cycle duration and rotational period $\log (\langle P_{\rm cyc} \rangle / P_{\rm rot})^2$ can be an observational measure of the dynamo number. Therefore, we shall compute the correlations of the stellar cycle variabilities with three key parameters: the rotational period $P_{\rm rot}$, the inverse squared Rossby number $(Ro)^{-2}$ and $\log (\langle P_{\rm cyc} \rangle / P_{\rm rot})^2$.

We start our analysis with the GK19 stars, which were selected from the MWO dataset based on their good quality and unambiguous cycles. Then, we study the GK25 set of stars.
We first only consider the base sets, and then add the Sun to the set, to bring the solar cycles in context.
Further, since we are able to fit cycles of all GK19 and GK25 stars with the quasi-Planck function, we also examine the correlation trends with these fitted data.
Finally, we evaluate the larger set of B95 stars, first without the solar data and then with the Sun included.
As described in \Sec{sec:mwodata}, the data of additional B95 stars cannot be fitted with the quasi-Planck function, so we only study the correlations with raw data.
For each pair of variability and rotation-related parameters, we calculate the Spearman rank correlation coefficient between the variables. 
A summary of the results thus obtained for all the data cases considered is presented in \Tab{tab:corr}.
Let us now discuss the most significant of these results in detail.


\subsubsection{GK19 and GK25 stars}

The only statistically significant correlations we find for the GK19 and GK25 set of stars are the following:
\begin{enumerate}[nosep]
    \item GK19 raw data with $(Ro)^{-2}$ at $-0.53$ ($p=0.013$) [\Fig{fig:gk19rossby}]
    \item GK19 quasi-Planck fit data including the Sun, with $P_{\rm rot}$ at $0.53$ ($0.011$)
    \item GK25 quasi-Planck fit data with $P_{\rm rot}$ at $0.53$ ($0.004$)
    \item GK25 quasi-Planck fit data including the Sun, with $P_{\rm rot}$ at $0.52$ ($0.004$) [\Fig{fig:gk25p+prot}]
\end{enumerate}
These four correlations are weak, but have $p\le0.015$ (GK25 sets have $p\le0.010$).
In spite of this, (2)--(4) seem to suggest that slowly rotating stars tend to have higher variability in their magnetic cycles, while (1) points to a similar trend for stars with large Rossby number.

\begin{figure}[t]
    \centering
    \includegraphics[width=\linewidth]{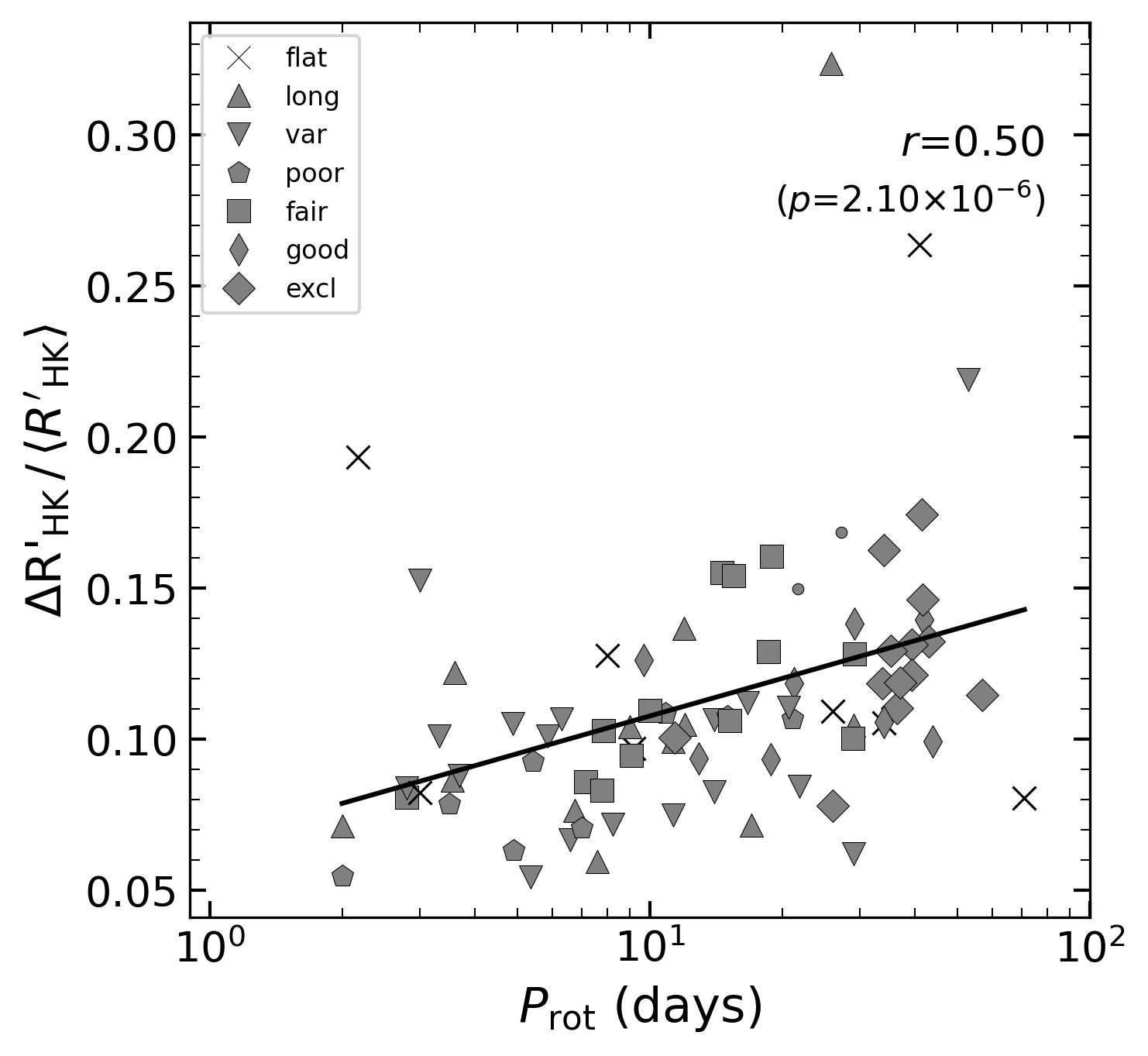}
    \caption{%
    Scatter plot between the parametrized variability and the rotational period for the B95$+$ set of stars. The variability class data is obtained from Table 5.3 in \cite{Egeland:2017:thesis}. Stars are classified into different variability classes, either non-cyclic: `flat', `long' (may have $P_{\rm cyc}>20$ yrs) and `var' (highly variable); or cyclic: `poor', `fair', `good' and `excellent'. Stars with the small dot marker do not have a variability class. Rest of the layout remains same as \Fig{fig:gk19rossby}.}
\label{fig:b95+prot}
\end{figure}

All other correlations are poor and are not statistically significant (\Tab{tab:corr}). Inclusion of the Sun improves the $P_{\rm rot}$ and $(\langle P_{\rm cyc}\rangle / P_{\rm rot})^2$ correlations, but worsens the correlations with $(Ro)^{-2}$. This may be because the Sun has a relatively large Rossby number, while its variability is particularly small in comparison to other stars in the study set.

Further, we should remember that the cycle period that is used here is computed based on a short time series, having only a few cycles. Therefore, we lack sufficient good quality data to conclusively say whether the changes in the stellar variability follow any uniform trend with respect to $(\langle P_{\rm cyc}\rangle / P_{\rm rot})^2$ and whether or not this parameter is a good measure of the stellar dynamo number.

Lastly, the seven additional stars in GK25 set have ill-defined cyclic behaviour as compared to the GK19 stars. This is why the correlations are not much different between the two sets, apart from those for $(\langle P_{\rm cyc}\rangle / P_{\rm rot})^2$, which improve slightly due to a few more values available. It is also generally seen that fitting the cycles with the quasi-Planck function improves or maintains the correlations in all cases, in comparison to its raw data counterpart. This suggests that the quasi-Planck smoothing enhances signal coherence in the variability-rotation period relations by suppressing short-term stochastic variability, thereby reinforcing the underlying physical trends.


\subsubsection{B95 stars}

\begin{figure}[t]
    \centering
    \includegraphics[width=\linewidth]{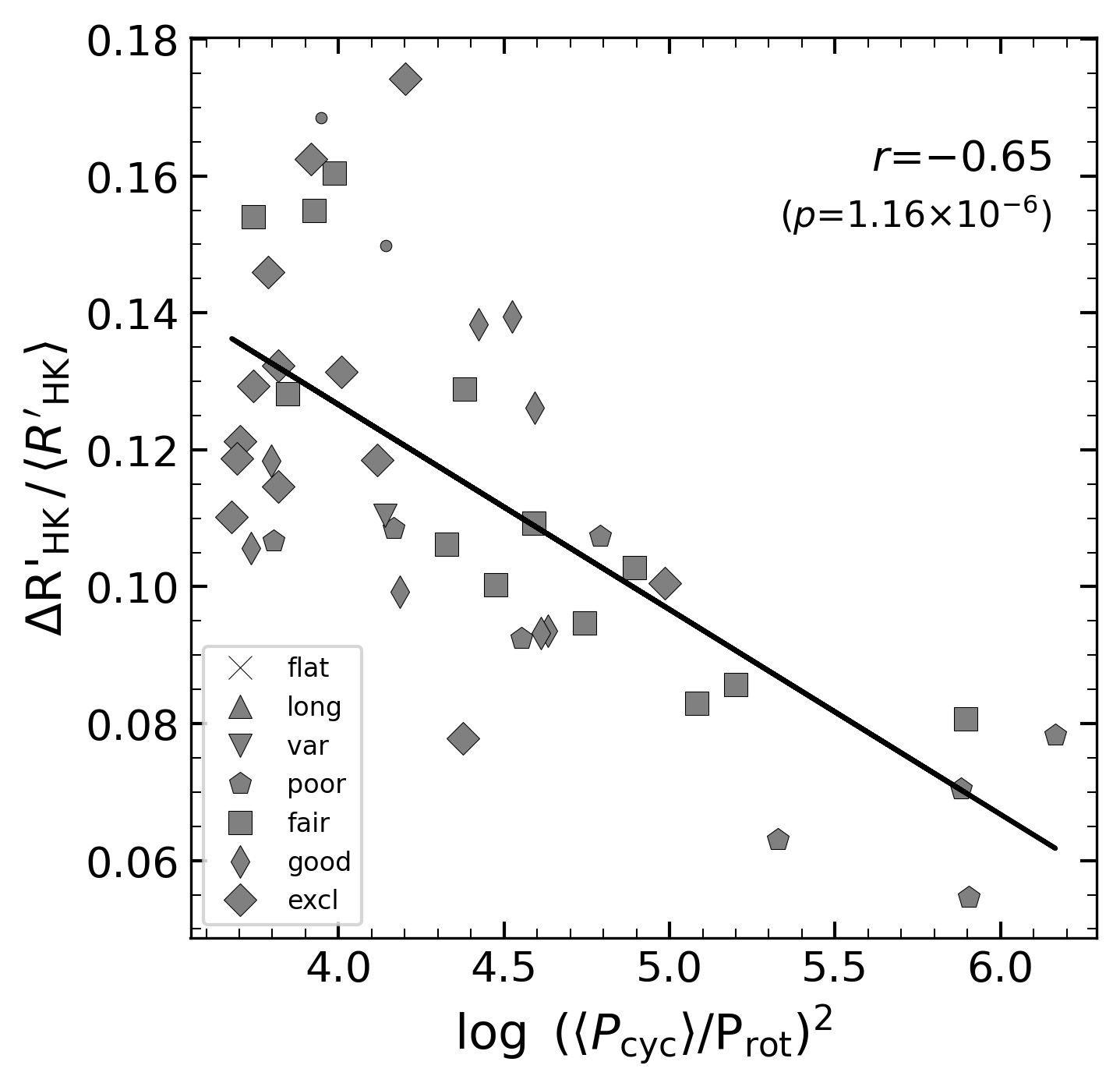}
    \caption{Same as \Fig{fig:b95+prot} but for evaluating the variability and $P_{\rm cyc}/P_{\rm rot}$ correlations. Only the stars that have the mean cycle period available in \Tab{tab:data} are included.}
\label{fig:b95+pcyc}
\end{figure}

Finally, we use the B95 set of stars to study the correlations between the stellar variability and the three rotation-related parameters.
We find that all the B95 and B95$+$ correlations are statistically significant and have $p\le0.010$ (except for one value which has $p = 0.012$).
This is likely due to inclusion of much larger number of stars in this set as compared to GK19 and GK25 sets. However, these correlations are still modest at best.
The Spearman coefficients are: $0.51$ ($p= 9.38 \times 10^{-7}$),\, $-0.30$ ($0.007$) and $-0.66$ ($1.05 \times 10^{-6}$)\, for correlations with $P_{\rm rot}$, $(Ro)^{-2}$ and $(\langle P_{\rm cyc}\rangle / P_{\rm rot})^2$ respectively; while for B95$+$ set they are: $0.50$ ($2.1 \times 10^{-6}$) [\Fig{fig:b95+prot}], $-0.28$ ($0.012$) and $-0.65$ ($1.16 \times 10^{-6}$) [\Fig{fig:b95+pcyc}] respectively.
Notably, the $P_{\rm rot}$ and $(Ro)^{-2}$ correlations remain consistent with the GK19 and GK25 stars, whereas the $(\langle P_{\rm cyc}\rangle / P_{\rm rot})^2$ are much improved in comparison. Improvement in the $(\langle P_{\rm cyc}\rangle / P_{\rm rot})^2$ correlations could be attributed to larger sample size of the set. The additional stars in B95 set have improper cycle variability and have much shorter rotation period than the solar analogue GK19 stars. Consistent, statistically significant results for $P_{\rm rot}$ again suggest that fast rotators tend to have less variable magnetic cycles.


\subsection{Variability in the modelled data}
\label{sec:res-mod}

\begin{figure}[ht]
\centering
\includegraphics[width=\linewidth]{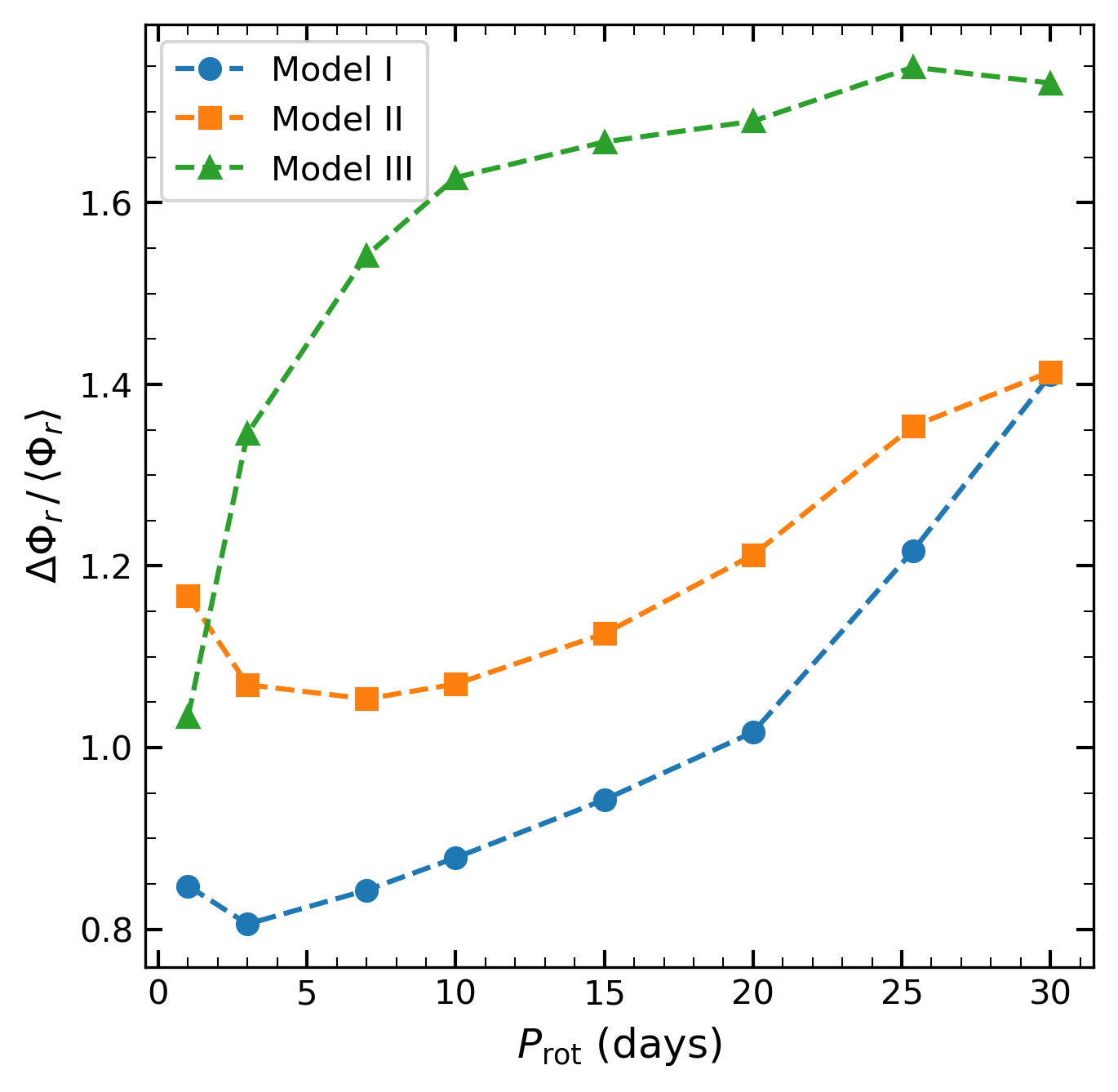}
\caption{Stellar cycle variability as obtained from three different dynamo models as a function of the stellar rotation period. The dynamo models are discussed in detail in \Sec{sec:moddata}, while the variability is defined in \Eq{eq:modvariability}.}
\label{fig:mod-var}
\end{figure}

\begin{figure}[ht]
\centering
\includegraphics[width=\linewidth]{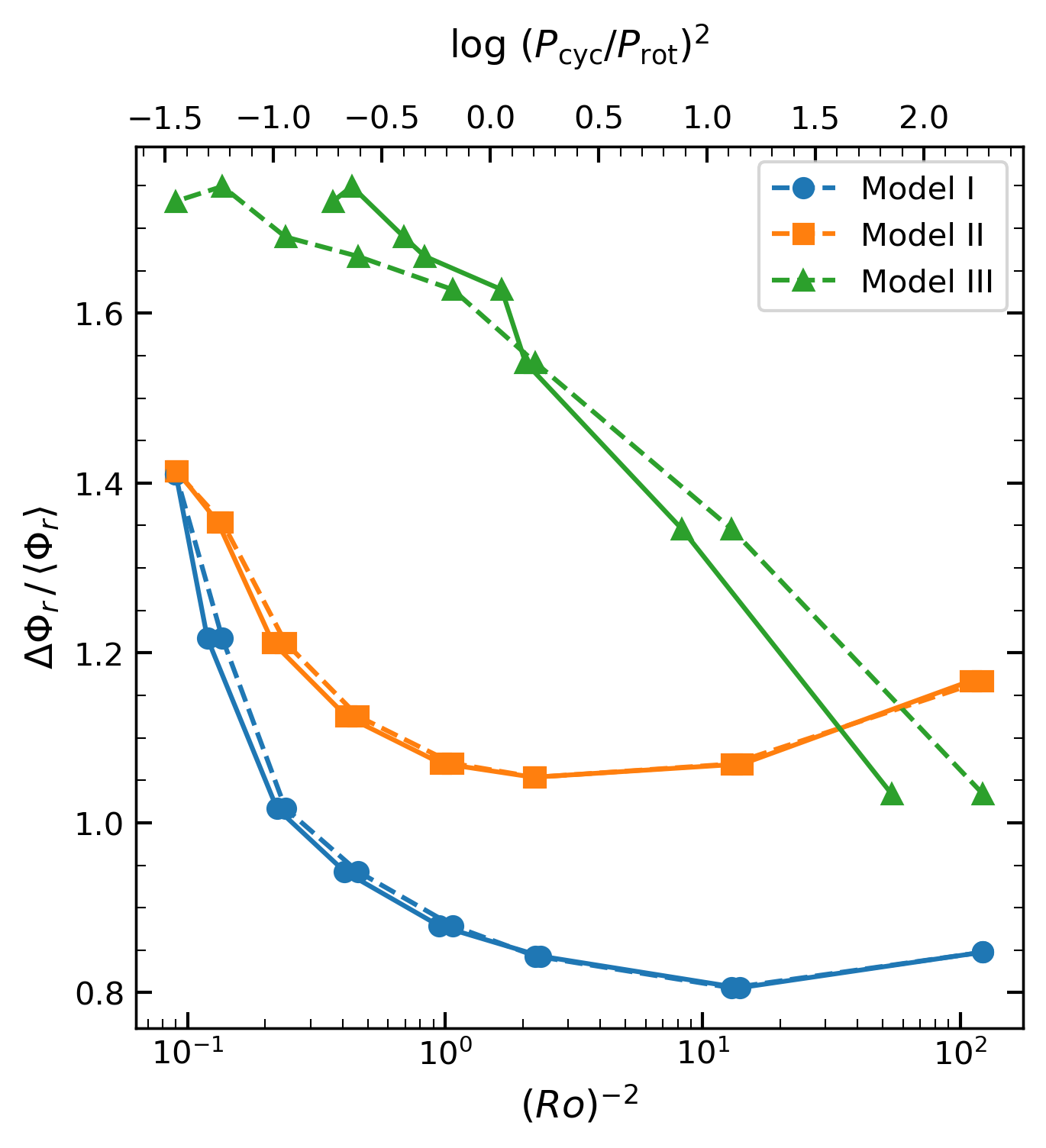}
\caption{Stellar cycle variability as obtained from three different dynamo models as a function of the Rossby number (dashed line and bottom x-axis) and $\log \, (P_{\rm cyc} / P_{\rm rot})^2$ (solid line and top x-axis). Models and the variability definition are the same as in \Fig{fig:mod-var}.}
\label{fig:mod-var2}
\end{figure}

Now, we come to the discussion about cycle variability seen for our dynamo model data. \Fig{fig:mod-var} shows the variability as a function of the rotation period. While different models show somewhat different variation, all show more or less increasing trend, except for a few rapidly rotating stars (that have $P_{\rm rot} < 5$~days) in Models~I and II, for which the variability first decreases and then increases.  We recall that in our dynamo models, we have considered all the stars of similar type with the same mass of $1M_\odot$, and thus our stellar model data represents only a small sample of the observed stars, which have varying structures.
Nevertheless, our simulations are still helpful in comparing them with the observations, particularly in terms of how the magnetic variability relates to the dynamo activity.  Comparing the model variability with that found in observational data (see \Figs{fig:gk25p+prot}{fig:b95+prot}), we can conclude that the model behaves in a similar manner to that of the observations. We also find that the stellar magnetic field variability decreases with the increase of $Ro^{-2}$ and $P_{\rm cyc}/P_{\rm rot}$ as shown in \Fig{fig:mod-var2}.

Following the predictions of \citet{DL77} and \citet{SBZ93} that these two quantities respectively are the theoretical and observed measures of the dynamo number ($D$), we compute D of our model stars using the following definition: 
\begin{equation}
    D =\frac{\alpha_0 \nabla \Omega L^3  } {\eta_0^2},
\end{equation}
where $\alpha_0$ is the strength of the \bl\ $\alpha$ parameter, $\nabla \Omega $ is the amount of the latitudinal shear in CZ, $L$ is the depth of CZ, and $\eta_0$ is the value of the turbulent diffusivity in CZ.  Fitting the data of the model III stars for the dynamo number and the Rossby number, we find that $D$ fits well with $Ro$ in the following form: $D \propto Ro^{-0.6}$, instead of $Ro^{-2}$ as found in the linear $\alpha \Omega$ dynamo model by \cite{DL77}.
Weaker dependence of $D$ on $Ro$ is possibly because of the decrease of the shear with the increase of rotation rate \citep{KR99, KO12b}, instead of the linear increase as assumed in the theory \citep{DL77}.
Further, doing the same with $(\langle P_{\rm cyc} \rangle / P_{\rm rot})$, we find that $D$ also scales with $(\langle P_{\rm cyc} \rangle / P_{\rm rot})^{0.6}$, in contrast to $\log \, (\langle P_{\rm cyc} \rangle / P_{\rm rot})^2$ as suggested by \citet{SBZ93}.
Nevertheless, these results give us confidence for interpreting the dynamo number using proxies in the form of $Ro$ or $P_{\rm cyc}/P_{\rm rot}$, which are available for a number of stars.


\section{Conclusion}
\label{sec:conclu}
In this work, we have studied the stellar cycle variability for stars selected from the MWO dataset. 
Our key contributions are:
\begin{itemize}[nosep]
    \item Inclusion of full 37 year $S$-index data, ranging 1966--2003, in the study of different sets of MWO stars with simple solar-like to complex aperiodic cycles (\Tab{tab:data}).
    \item Stellar variability computation in terms of $R^{\prime}_{\rm HK}$ from the $S$-index data using \Eq{eq:var-rhk}.
    \item Exploration of how rotation affects the stellar variability using correlations with three rotation-related parameters, $P_{\rm rot}$, $(Ro)^{-2}$ and $\log\,(\langle P_{\rm cyc} \rangle / P_{\rm rot})^2$.
    \item Constraining theoretical dynamo models using the MWO observation data and analysis.
\end{itemize}

We find that the stellar cycle variability decreases with increase of rotation rate (or decrease of rotation period). Similar trends are also detected with the inverse squared Rossby number and $(\langle P_{\rm cyc} \rangle / P_{\rm rot})^2$.
Across all our GK19 and GK25 sets, the best significant correlations are for $P_{\rm rot}$ and $(Ro)^{-2}$ respectively at $0.53$ and $-0.53$.
On the other hand, the correlations with $(\langle P_{\rm cyc} \rangle / P_{\rm rot})^2$ are weak and have low significance ($p>0.01$). This is mainly due to the requirement for good quality data for several cycles. Limited availability of data (only a few usable cycles available) is a major hindrance in probing the true nature of the trend in this case.

In contrast, for the B95 sets, all the correlations are statistically significant with $p\le0.01$. This is likely because of increase in the number of stars included, as compared to GK19 and GK25. These B95 and B95$+$ set correlations are, however, still modest at best. The best correlation among these set of stars is found at $-0.66$ for $(\langle P_{\rm cyc} \rangle / P_{\rm rot})^2$. Nevertheless, the correlation trends obtained across different rotation-related parameters are consistent and point to an indication that slowly rotating stars have highly varying stellar activity.

Furthermore, from independent explorations using data from three different dynamo models \citep{VKK23} (though with certain limitations, namely, same mass for all stars and uncertainties in the dynamo parameters), we find a qualitatively similar trend in the stellar magnetic variability.
Following the work of \citet{DL77} and \citet{SBZ93}, we also checked how  the quantities: $(Ro)^{-2}$ and $\log \, (\langle P_{\rm cyc} \rangle / P_{\rm rot})^2$, are related to the dynamo number.
We find that the model data fits well for the Dynamo number $D$ scaling with $(Ro)^{-0.6}$, instead of $(Ro)^{-2}$ as suggested in the linear dynamo theory.
This weaker dependence may possibly be because of the shear decreasing with the increase of rotation rate \citep{KR99, KO12b}, instead of the linear increase as assumed in the theory \citep{DL77}. Further, we also find that the data fits well for $D$ scaling with $(\langle P_{\rm cyc} \rangle / P_{\rm rot})^{0.6}$, in contrast to $\log \, (\langle P_{\rm cyc} \rangle / P_{\rm rot})^2$ as suggested by \citet{SBZ93}.

In summary, the variability decreases with the increase of rotation rate or the dynamo number.  These results nicely align with the expectation from the dynamo model, which suggests that in the highly supercritical regime, due to non-linear quenching, the growth of the magnetic field is limited, and the magnetic field does not vary appreciably. Weaker variability at high rotation rate or dynamo number also suggests that, in the parameter regimes of the MWO stellar data, the dynamo may not operate in chaotic regime. 




\appendix
\label{sec:appendix}
In \Tab{tab:data} we provide a list of properties for all the 81 MWO stars (and the Sun) used in our study. The correlations in \Tab{tab:corr} and the scatter plots in \Sec{sec:result}, can be reproduced using these data. \Tab{tab:fitParams} tabulates the fit parameters and cycle properties for the additional 7 stars in GK25 set.


\startlongtable
\begin{deluxetable}{lccclccc}
\tabletypesize{\scriptsize}
\tablewidth{0pt}        
\tablecaption{Properties of all the stars considered in this analysis.
\label{tab:data}}
\tablehead{
\colhead{\textbf{Star}} &
\colhead{$b_v$} &
\colhead{$P_{\rm rot}$} &
\colhead{$Ro$} &
\colhead{\#cyc} &
\colhead{$\langle P_{\rm cyc}\rangle$} &
\colhead{var($R$)} &
\colhead{$\Delta R^{\prime}_{\rm HK}$}
}
\startdata
Sun & 0.66 & 27.27 & 2.15 & 9+1 & 10.63 & 0.077 & 8.780e-07 \\
\midrule
\multicolumn{8}{c}{\textbf{GK19 stars}}\\
\midrule
HD 3651   & 0.85 & 44.00 & 1.93 & 2+1 & 14.91 & 0.099 & 9.629e-07 \\
HD 4628   & 0.88 & 41.60 & 1.79 & 3+2 & 8.91  & 0.146 & 2.033e-06 \\
HD 10476  & 0.84 & 33.70 & 1.71 & 3+1 & 10.56 & 0.119 & 1.374e-06 \\
HD 16160  & 0.98 & 57.00 & 2.10 & 2+1 & 12.65 & 0.115 & 1.307e-06 \\
HD 20630  & 0.68 & 9.08  & 0.67 & 3+2 & 5.85  & 0.095 & 3.473e-06 \\
HD 26965  & 0.82 & 43.00 & 2.15 & 3+1 & 9.54  & 0.132 & 1.718e-06 \\
HD 32147  & 1.06 & 39.30 & 2.05 & 2+2 & 10.87 & 0.131 & 1.449e-06 \\
HD 81809  & 0.64 & 39.30 & 3.52 & 4+1 & 7.64  & 0.121 & 1.471e-06 \\
HD 103095 & 0.75 & 36.50 & 1.79 & 4+1 & 6.89  & 0.110 & 1.385e-06 \\
HD 115404 & 0.93 & 18.80 & 0.83 & 2+2 & 10.42 & 0.093 & 2.988e-06 \\
HD 149661 & 0.81 & 21.30 & 1.05 & 7   & 4.61  & 0.119 & 3.023e-06 \\
HD 152391 & 0.76 & 11.40 & 0.65 & 3+1 & 9.70  & 0.101 & 3.545e-06 \\
HD 155885 & 0.86 & 21.11 & 1.00 & 4+2 & 4.60  & 0.107 & 3.009e-06 \\
HD 155886 & 0.86 & 20.69 & 0.98 & 2+3 & 6.67  & 0.110 & 2.890e-06 \\
HD 156026 & 1.16 & 29.20 & 0.87 & 2+1 & 13.02 & 0.138 & 3.168e-06 \\
HD 160346 & 0.96 & 35.30 & 1.61 & 4+1 & 7.19  & 0.129 & 2.004e-06 \\
HD 161239 & 0.65 & 29.20 & 2.44 & 5   & 6.69  & 0.128 & 8.463e-07 \\
HD 166620 & 0.87 & 41.50 & 1.99 & 2+1 & 14.34 & 0.174 & 1.934e-06 \\
HD 201091 & 1.18 & 37.10 & 1.45 & 4+1 & 7.13  & 0.119 & 2.024e-06 \\
HD 201092 & 1.37 & 34.10 & 1.46 & 3+2 & 6.89  & 0.106 & 1.340e-06 \\
HD 219834B& 0.91 & 34.00 & 1.95 & 4   & 8.45  & 0.162 & 1.877e-06 \\
\midrule
\multicolumn{8}{c}{\textbf{Additional GK25 stars}}\\
\midrule
HD 1835    & 0.66 & 7.84 & 0.62 & 5+1  & 6.00 & 0.103 & 3.763e-06 \\
HD 76151   & 0.67 & 15.2 & 1.56 & 6  & 6.06 & 0.106 & 2.204e-06 \\
HD 78366   & 0.6  & 9.7  & 1.06 & 7  & 5.26 & 0.126 & 3.188e-06 \\
HD 100180  & 0.57 & 14.6 & 1.88 & 9  & 3.67 & 0.155 & 1.872e-06 \\
HD 114710  & 0.57 & 12.9 & 1.66 & 4+2  & 7.31 & 0.094 & 1.713e-06 \\
HD 165341A & 0.86 & 18.9 & 0.95 & 6+1  & 5.10 & 0.160 & 4.452e-06 \\
HD 190406  & 0.61 & 15.5 & 1.44 & 11 & 3.19 & 0.154 & 2.514e-06 \\
\midrule
\multicolumn{8}{c}{\textbf{Other B95 stars}}\\
\midrule
HD 2454    & 0.43 & 3.0   & 1.54 & - & -     & 0.153 & 2.478e-06 \\
HD 3229    & 0.44 & 2.0   & 0.91 & - & 4.90  & 0.055 & 1.416e-06 \\
HD 6920    & 0.6  & 14.0  & 1.54 & - & -     & 0.083 & 1.313e-06 \\
HD 9562    & 0.64 & 29.0  & 2.54 & - & -     & 0.105 & 7.098e-07 \\
HD 10700   & 0.72 & 34.0  & 2.14 & - & -     & 0.105 & 1.162e-06 \\
HD 10780   & 0.81 & 21.7  & 1.10 & - & 7.00  & 0.150 & 3.074e-06 \\
HD 12235   & 0.62 & 14.0  & 1.37 & - & -     & 0.106 & 1.061e-06 \\
HD 13421   & 0.56 & 2.173 & 0.31 & - & -     & 0.193 & 1.193e-06 \\
HD 16673   & 0.52 & 6.6   & 1.32 & - & -     & 0.067 & 1.465e-06 \\
HD 17925   & 0.87 & 6.76  & 0.32 & - & -     & 0.076 & 3.670e-06 \\
HD 18256   & 0.43 & 2.8   & 1.44 & - & 6.80  & 0.081 & 1.505e-06 \\
HD 22049   & 0.88 & 11.3  & 0.52 & - & -     & 0.099 & 3.553e-06 \\
HD 23249   & 0.92 & 71.0  & 3.20 & - & -     & 0.080 & 5.296e-07 \\
HD 25998   & 0.46 & 2.0   & 0.72 & - & -     & 0.071 & 2.635e-06 \\
HD 26913   & 0.7  & 7.15  & 0.48 & - & 7.80  & 0.086 & 3.286e-06 \\
HD 30495   & 0.63 & 11.3  & 1.04 & - & -     & 0.075 & 2.278e-06 \\
HD 35296   & 0.53 & 3.56  & 0.65 & - & -     & 0.086 & 3.400e-06 \\
HD 37394   & 0.84 & 10.86 & 0.53 & - & 3.60  & 0.108 & 3.777e-06 \\
HD 39587   & 0.59 & 5.36  & 0.63 & - & -     & 0.054 & 1.983e-06 \\
HD 45067   & 0.56 & 8.0   & 1.16 & - & -     & 0.128 & 1.052e-06 \\
HD 72905   & 0.62 & 4.89  & 0.48 & - & -     & 0.105 & 4.342e-06 \\
HD 75332   & 0.49 & 3.7   & 0.98 & - & -     & 0.088 & 2.998e-06 \\
HD 82443   & 0.77 & 5.42  & 0.30 & - & 2.80  & 0.092 & 5.732e-06 \\
HD 82885   & 0.77 & 18.6  & 1.02 & - & 7.90  & 0.129 & 3.016e-06 \\
HD 89744   & 0.54 & 9.2   & 1.55 & - & -     & 0.097 & 7.181e-07 \\
HD 95735   & 1.51 & 53.0  & 1.95 & - & -     & 0.219 & 8.201e-07 \\
HD 97334   & 0.61 & 8.25  & 0.85 & - & -     & 0.072 & 2.687e-06 \\
HD 101501  & 0.72 & 16.68 & 1.05 & - & -     & 0.112 & 3.072e-06 \\
HD 107213  & 0.5  & 9.0   & 2.16 & - & -     & 0.104 & 7.891e-07 \\
HD 115043  & 0.6  & 5.861 & 0.64 & - & -     & 0.101 & 3.683e-06 \\
HD 115383  & 0.58 & 3.33  & 0.42 & - & -     & 0.101 & 3.659e-06 \\
HD 115617  & 0.71 & 29.0  & 1.89 & - & -     & 0.062 & 6.142e-07 \\
HD 120136  & 0.47 & 3.5   & 1.14 & - & 11.60 & 0.078 & 1.426e-06 \\
HD 124570  & 0.54 & 26.0  & 4.39 & - & -     & 0.109 & 7.556e-07 \\
HD 124850  & 0.52 & 7.6   & 1.52 & - & -     & 0.059 & 1.194e-06 \\
HD 129333  & 0.61 & 2.8   & 0.29 & - & -     & 0.084 & 5.895e-06 \\
HD 131156B & 1.17 & 11.94 & 0.49 & - & -     & 0.137 & 4.833e-06 \\
HD 131156A & 0.76 & 6.31  & 0.35 & - & -     & 0.107 & 4.606e-06 \\
HD 141004  & 0.6  & 25.8  & 2.84 & - & -     & 0.324 & 3.283e-06 \\
HD 143761  & 0.6  & 17.0  & 1.87 & - & -     & 0.071 & 6.420e-07 \\
HD 154417  & 0.57 & 7.78  & 1.04 & - & 7.40  & 0.083 & 2.413e-06 \\
HD 176095  & 0.46 & 3.6   & 1.30 & - & -     & 0.122 & 2.482e-06 \\
HD 182572  & 0.77 & 41.0  & 2.25 & - & -     & 0.264 & 2.176e-06 \\
HD 185144  & 0.8  & 27.2  & 1.40 & - & 7.00  & 0.169 & 2.545e-06 \\
HD 187691  & 0.55 & 10.0  & 1.56 & - & 5.40  & 0.109 & 1.028e-06 \\
HD 190007  & 1.17 & 28.95 & 1.19 & - & 13.70 & 0.100 & 1.968e-06 \\
HD 194012  & 0.51 & 7.0   & 1.53 & - & 16.70 & 0.070 & 1.356e-06 \\
HD 206860  & 0.59 & 4.91  & 0.58 & - & 6.20  & 0.063 & 2.397e-06 \\
HD 207978  & 0.42 & 3.0   & 1.74 & - & -     & 0.082 & 1.061e-06 \\
HD 212754  & 0.52 & 12.0  & 2.40 & - & -     & 0.105 & 9.026e-07 \\
HD 217014  & 0.67 & 21.9  & 1.67 & - & -     & 0.084 & 7.087e-07 \\
HD 219834A & 0.8  & 42.0  & 2.17 & - & 21.00 & 0.139 & 1.199e-06 \\
HD 224930  & 0.67 & 15.01 & 1.14 & - & 10.20 & 0.107 & 1.422e-06 \\
\enddata
\tablecomments{GK19 stars were also studied by \cite{Garg19}, while the full set was studied in \cite{Baliu95}.
$b_v$ is the B$-$V color value of the stars taken from \cite{SBZ94} if available, otherwise from \cite{B07} or from \cite{Noyes84a}.
$P_{\rm rot}$ is the rotational period of the star in days, taken from \cite{Ol16} for GK19 and the additional GK25 stars, and from \cite{Egeland:2017:thesis} for the other B95 stars.
The column $Ro$ denotes the Rossby number, taken from \cite{Egeland:2017:thesis}. ``\#cyc" denotes the number of usable full+partial cycles for each stars when doing a cycle fitting.
$\langle P_{\rm cyc} \rangle$ is the mean cycle period in years. It is computed using the cycle period of all full cycles for the GK19 and GK25 stars. For the other B95 stars, the values are from Table 5.3 of \cite{Egeland:2017:thesis}.
var($R$) is the parametrized variability computed by \Eq{eq:var-rhk} and $\Delta R^{\prime}_{\rm HK}$ denotes the standard deviation in $R^{\prime}_{\rm HK}$.
}
\end{deluxetable}

\begin{deluxetable*}{l*{8}{c}}
\tablecaption{Cycle fit parameters for the additional GK25 stars.
\label{tab:fitParams}}
\tablehead{
\colhead{Star} &
\colhead{cyc} &
\colhead{$t_{\rm start}$} &
\colhead{$P_{\rm cyc}$} &
\colhead{$a$} &
\colhead{$b$} &
\colhead{$c$} &
\colhead{$t_0$} &
\colhead{$\chi^2_{\rm red}$}
}
\startdata
HD 1835  & 1* & 1967.805 & -     & 0.000001      & 261.22799      & 1.11764       & 1843.470 & 11.6 \\
         & 2  & 1972.638 & 4.67  & $-$2.293e+05   & $-$11.17197      & $-$8.342e+10    & 2027.498 & 14.9 \\
         & 3  & 1977.305 & 4.5   & 0.00039      & 11.94459       & 0.84901       & 1967.274 & 10.9 \\
         & 4  & 1981.805 & 6.58  & 0.00011      & 17.99733       & 0.77755       & 1964.655 & 18.4 \\
         & 5  & 1988.388 & 8.83  & 0.00009      & 19.10546       & 0.88719       & 1975.030 & 63.0 \\
         & 6  & 1997.222 & 5.42  & 0.00073      & 9.96229        & 0.84232       & 1989.521 & 18.2 \\
         & 7* & 2002.638 & -     & $-$0.03676     & 4.20228        & 1.00233       & 2002.362 & 1.5 \\
\hline
HD 76151  & 1  & 1966.600 & 3.15  & 6.112e+27    & $-$7.31098       & $-$5.478e+33    & 1905.591 & 4.8 \\
          & 2  & 1969.750 & 7.99  & 3.669e+70    & 12.79014       & $-$6.604e+77    & 1806.756 & 18.5 \\
          & 3  & 1977.742 & 5.01  & 2165.40983   & 11.54279       & $-$8.946e+08    & 1930.954 & 13.5 \\
          & 4  & 1982.750 & 6.32  & 0.00003      & 25.86506       & 0.55687       & 1953.410 & 12.7 \\
          & 5  & 1989.075 & 7.92  & 8.671e+22    & 13.41582       & $-$4.238e+29    & 1887.245 & 15.8 \\
          & 6  & 1997.000 & 5.98  & 1.348e+19    & 18.58360       & $-$1.476e+26    & 1859.945 & 16.9 \\
\hline
HD 78366  & 1  & 1966.189 & 5.23  & 0.00055      & 9.66249        & 0.83731       & 1958.288 & 7.2 \\
          & 2  & 1971.417 & 5.25  & 11.47147     & 9.56138        & $-$1.473e+06    & 1941.336 & 21.5 \\
          & 3  & 1976.667 & 7.0   & 0.00023      & 13.22786       & 0.70549       & 1965.499 & 15.7 \\
          & 4  & 1983.667 & 5.58  & 1.388e+70    & 9.33569        & $-$1.012e+77    & 1865.494 & 14.0 \\
          & 5  & 1989.250 & 3.5   & 0.00125      & 7.42016        & 0.80933       & 1983.334 & 11.3 \\
          & 6  & 1992.750 & 5.33  & 0.00084      & 8.50172        & 0.83849       & 1986.630 & 25.2 \\
          & 7  & 1998.084 & 4.9   & $-$1.954e+47   & $-$8.40719       & $-$5.288e+53    & 2091.084 & 20.7 \\
\hline
HD 100180 & 1  & 1969.326 & 2.83  & 0.00085      & 7.20387        & 0.80984       & 1962.922 & 3.3 \\
          & 2  & 1972.159 & 4.17  & $-$0.00004     & $-$19.19486      & 0.81957       & 1994.468 & 4.4 \\
          & 3  & 1976.326 & 3.17  & 1.619e+05    & 7.78235        & $-$4.244e+10    & 1941.825 & 10.7 \\
          & 4  & 1979.492 & 3.67  & 0.04391      & 11.30797       & $-$5678.51508   & 1950.765 & 7.0 \\
          & 5  & 1983.159 & 3.59  & 3.397e+05    & 15.57806       & $-$8.894e+11    & 1907.431 & 5.6 \\
          & 6  & 1986.750 & 3.85  & 1.466e+57    & $-$6.00941       & $-$3.001e+63    & 1918.016 & 34.5 \\
          & 7  & 1990.600 & 4.31  & 0.00041      & 8.64593        & 1.07565       & 1985.289 & 5.9 \\
          & 8  & 1994.909 & 2.25  & 0.00121      & 6.36164        & 0.83399       & 1989.520 & 5.6 \\
          & 9  & 1997.159 & 5.24  & 0.00003      & 23.75888       & 0.62241       & 1974.534 & 10.9 \\
\hline
HD 114710 & 1* & 1967.167 & -     & $-$0.00001     & 42.11102       & 0.93546       & 2000.995 & 6.3 \\
          & 2  & 1971.000 & 7.92  & 0.00010      & 14.72174       & 1.08196       & 1963.020 & 6.3 \\
          & 3  & 1978.917 & 9.67  & 0.00008      & 16.24180       & 1.07107       & 1969.454 & 8.0 \\
          & 4  & 1988.584 & 5.33  & 0.00300      & 14.76241       & $-$388.77329    & 1957.939 & 4.5 \\
          & 5  & 1993.917 & 6.33  & 0.00009      & 16.13428       & 0.84125       & 1980.601 & 13.3 \\
          & 6* & 2000.250 & -     & 0.00036      & 10.34536       & 0.92631       & 1992.986 & 13.0 \\
\hline
HD 165341A & 1 & 1967.485 & 5.04  & 0.00337      & 6.40174        & 0.82060       & 1963.267 & 30.4 \\
           & 2 & 1972.520 & 5.75  & 1.480e+108   & 49.91386       & $-$1.970e+117   & 1159.580 & 32.2 \\
           & 3 & 1978.270 & 4.08  & $-$0.07014     & 7.00069        & $-$520.52153    & 1996.586 & 19.3 \\
           & 4 & 1982.354 & 5.92  & $-$0.00203     & 7.94594        & 0.18972       & 1994.831 & 56.0 \\
           & 5 & 1988.270 & 5.73  & 0.00293      & 6.53915        & 0.90057       & 1984.283 & 28.7 \\
           & 6* & 1994.000 & -    & 0.00263      & 7.26589        & 0.05043       & 1986.054 & 21.5 \\
           & 7 & 1998.104 & 4.1   & 0.00794      & 4.72279        & 0.78545       & 1994.987 & 150.3 \\
\hline
HD 190406 & 1  & 1968.000 & 2.64  & 2.934e+23    & 5.86217        & $-$1.458e+29    & 1923.176 & 4.6 \\
          & 2  & 1970.639 & 2.75  & 0.00418      & 4.22540        & 1.06836       & 1968.775 & 3.2 \\
          & 3  & 1973.389 & 2.25  & 0.01060      & 3.11472        & 1.06513       & 1972.105 & 3.5 \\
          & 4  & 1975.639 & 3.36  & 0.00415      & 4.24038        & 1.06712       & 1973.901 & 16.6 \\
          & 5  & 1979.000 & 2.5   & 8.318e+51    & 6.45312        & $-$1.663e+58    & 1907.385 & 10.4 \\
          & 6  & 1981.500 & 2.56  & 0.01510      & 4.08737        & $-$6.41336      & 1976.371 & 6.8 \\
          & 7  & 1984.056 & 5.42  & 0.00004      & 21.80507       & 0.69536       & 1960.578 & 9.0 \\
          & 8  & 1989.472 & 4.33  & 0.00053      & 8.36961        & 1.08400       & 1985.117 & 5.8 \\
          & 9  & 1993.806 & 2.89  & 0.00604      & $-$3.76621       & 1.06340       & 1992.336 & 41.6 \\
          & 10 & 1996.700 & 2.9   & 0.00763      & 3.51947        & 1.05529       & 1995.463 & 9.0 \\
          & 11 & 1999.600 & 3.06  & 0.00234      & 5.62410        & 0.79158       & 1994.900 & 17.0 \\
\enddata
\tablecomments{$t_{\rm start}$ is the cycle start time, whereas $P_{\rm cyc}$ is the cycle period. Cycles marked with * are partial cycles and do not count towards the mean cycle period. $(a,b,c,t_0)$ are the parameters of the quasi-Planck function given by \Eq{eq:planck}. $\chi^2_{\rm red}$ is the reduced chi-square value for the cycle fit. Note that if the cycle is well-define, the fit is better and then ideally $t_{\rm start}$ and the parameter $t_0$ are close to eachother.}
\end{deluxetable*}


\section*{Acknowledgements}
We would like to thank Vindya Vashishth for helping us in analysing the stellar dynamo simulation data.
We also thank the anonymous referee for the kind suggestions that helped in improving the paper.
Ricky Egeland provided the initial MWO $S$-index data used in this study. We thank Willie Soon for valuable suggestions and useful clarifications.
S.G. is grateful to IIT-BHU for a project fellowship (remote) during the summer of 2021. B.B.K. acknowledges the financial support from the 
Anusandhan National Research Foundation (ANRF) through the MATRIC program (file no. MTR/2023/000670).
More information on the MWO dataset and the data release is available at: \url{https://nso.edu/data/historical-data/mount-wilson-observatory-hk-project/}.
The dataset is derived from the Mount Wilson Observatory HK Project, which was supported by both public and private funds through the Carnegie Observatories, the Mount Wilson Institute, and the Harvard-Smithsonian Center for Astrophysics starting in 1966 and continuing for over 36 years. These data are the result of the dedicated work of O. Wilson, A. Vaughan, G. Preston, D. Duncan, S. Baliunas, and many others.

%

\bibliographystyle{yahapj}
\bibliography{variabilityPaper}  

@incollection{Gil80,
       author = {{Gilman}, P.~A.},
        title = "{Differential rotation in stars with convection zones}",
    booktitle = {IAU Colloq. 51: Stellar Turbulence},
         year = 1980,
       editor = {{Gray}, D.~F. and {Linsky}, J.~L.},
       volume = {114},
        pages = {19-37},
          doi = {10.1007/3-540-09737-6_7},
       adsurl = {https://ui.adsabs.harvard.edu/abs/1980LNP...114...19G}
}

@ARTICLE{B07,
       author = {{Barnes}, Sydney A.},
        title = "{Ages for Illustrative Field Stars Using Gyrochronology: Viability, Limitations, and Errors}",
      journal = {\apj},
         year = 2007,
        month = nov,
       volume = {669},
       number = {2},
        pages = {1167-1189},
          doi = {10.1086/519295},
archivePrefix = {arXiv},
       eprint = {0704.3068},
 primaryClass = {astro-ph},
       adsurl = {https://ui.adsabs.harvard.edu/abs/2007ApJ...669.1167B}
}

@ARTICLE{Baliu95,
   author = {{Baliunas}, S.~L. and {Donahue}, R.~A. and {Soon}, W.~H. and 
	{Horne}, J.~H. and {Frazer}, J. and {Woodard-Eklund}, L. and 
	{Bradford}, M. and {Rao}, L.~M. and {Wilson}, O.~C. and {Zhang}, Q. and 
	{Bennett}, W. and {Briggs}, J. and {Carroll}, S.~M. and {Duncan}, D.~K. and 
	{Figueroa}, D. and {Lanning}, H.~H. and {Misch}, T. and {Mueller}, J. and 
	{Noyes}, R.~W. and {Poppe}, D. and {Porter}, A.~C. and {Robinson}, C.~R. and 
	{Russell}, J. and {Shelton}, J.~C. and {Soyumer}, T. and {Vaughan}, A.~H. and 
	{Whitney}, J.~H.},
    title = "{Chromospheric variations in main-sequence stars}",
  journal = {\apj},
     year = 1995,
    month = jan,
   volume = 438,
    pages = {269-287},
      doi = {10.1086/175072},
   adsurl = {http://adsabs.harvard.edu/abs/1995ApJ...438..269B}
}

@ARTICLE{Biswas23,
       author = {{Biswas}, Akash and {Karak}, Bidya Binay and {Usoskin}, Ilya and {Weisshaar}, Eckhard},
        title = "{Long-Term Modulation of Solar Cycles}",
      journal = {\ssr},
         year = 2023,
        month = apr,
       volume = {219},
       number = {3},
          eid = {19},
        pages = {19},
          doi = {10.1007/s11214-023-00968-w},
archivePrefix = {arXiv},
       eprint = {2302.14845},
 primaryClass = {astro-ph.SR},
       adsurl = {https://ui.adsabs.harvard.edu/abs/2023SSRv..219...19B}
}

@ARTICLE{Baum22,
       author = {{Baum}, Anna C. and {Wright}, Jason T. and {Luhn}, Jacob K. and {Isaacson}, Howard},
        title = "{Five Decades of Chromospheric Activity in 59 Sun-like Stars and New Maunder Minimum Candidate HD 166620}",
      journal = {\aj},
         year = 2022,
        month = apr,
       volume = {163},
       number = {4},
          eid = {183},
        pages = {183},
          doi = {10.3847/1538-3881/ac5683},
archivePrefix = {arXiv},
       eprint = {2203.13376},
 primaryClass = {astro-ph.SR},
       adsurl = {https://ui.adsabs.harvard.edu/abs/2022AJ....163..183B}
}

@ARTICLE{luhn22,
       author = {{Luhn}, Jacob K. and {Wright}, Jason T. and {Henry}, Gregory W. and {Saar}, Steven H. and {Baum}, Anna C.},
        title = "{HD 166620: Portrait of a Star Entering a Grand Magnetic Minimum}",
      journal = {\apjl},
         year = 2022,
        month = sep,
       volume = {936},
       number = {2},
          eid = {L23},
        pages = {L23},
          doi = {10.3847/2041-8213/ac8b13},
archivePrefix = {arXiv},
       eprint = {2207.00612},
 primaryClass = {astro-ph.SR},
       adsurl = {https://ui.adsabs.harvard.edu/abs/2022ApJ...936L..23L}
}

@ARTICLE{BoroSaikia18,
   author = {{Boro Saikia}, S. and {Marvin}, C.~J. and {Jeffers}, S.~V. and
        {Reiners}, A. and {Cameron}, R. and {Marsden}, S.~C. and {Petit}, P. and
        {Warnecke}, J. and {Yadav}, A.~P.},
    title = "{Chromospheric activity catalogue of 4454 cool stars. Questioning the active branch of stellar activity cycles}",
  journal = {\aap},
archivePrefix = "arXiv",
   eprint = {1803.11123},
 primaryClass = "astro-ph.SR",
     year = 2018,
    month = aug,
   volume = 616,
      eid = {A108},
    pages = {A108},
      doi = {10.1051/0004-6361/201629518},
   adsurl = {http://adsabs.harvard.edu/abs/2018A%26A...616A.108B}
}

@ARTICLE{Cha20,
       author = {{Charbonneau}, Paul},
        title = "{Dynamo models of the solar cycle}",
      journal = {Living Reviews in Solar Physics},
         year = 2020,
        month = jun,
       volume = {17},
       number = {1},
          eid = {4},
        pages = {4},
          doi = {10.1007/s41116-020-00025-6},
       adsurl = {https://ui.adsabs.harvard.edu/abs/2020LRSP...17....4C}
}

@ARTICLE{CSZ05,
       author = {{Charbonneau}, Paul and {St-Jean}, C{\'e}dric and {Zacharias}, Pia},
        title = "{Fluctuations in Babcock-Leighton Dynamos. I. Period Doubling and Transition to Chaos}",
      journal = {\apj},
         year = 2005,
        month = jan,
       volume = {619},
       number = {1},
        pages = {613-622},
          doi = {10.1086/426385},
       adsurl = {https://ui.adsabs.harvard.edu/abs/2005ApJ...619..613C}
}

@ARTICLE{DL77,
       author = {{Durney}, B.~R. and {Latour}, J.},
        title = "{On the angular momentum loss of late-type stars}",
      journal = {Geophysical and Astrophysical Fluid Dynamics},
         year = 1977,
        month = jan,
       volume = {9},
       number = {1},
        pages = {241-255},
          doi = {10.1080/03091927708242330},
       adsurl = {https://ui.adsabs.harvard.edu/abs/1977GApFD...9..241D}
}

@article{Garg19,
  title = {Waldmeier {{Effect}} in {{Stellar Cycles}}},
  author = {Garg, Suyog and Karak, Bidya Binay and Egeland, Ricky and Soon, Willie and Baliunas, Sallie},
  year = {2019},
  month = dec,
  journal = {\apj},
  volume = {886},
  number = {2},
  pages = {132},
  issn = {0004-637X, 1538-4357},
  doi = {10.3847/1538-4357/ab4a17},
  urldate = {2024-11-29},
  langid = {english}
}

@article{hathaway1994,
  title = {The {{Shape}} of the {{Sunspot Cycle}}},
  author = {Hathaway, David H. and Wilson, Robert M. and Reichmann, Edwin J.},
  year = {1994},
  month = apr,
  journal = {Solar Physics},
  volume = {151},
  pages = {177--190},
  issn = {0038-0938},
  doi = {10.1007/BF00654090},
  urldate = {2025-06-08},
}

@ARTICLE{Hazra19,
       author = {{Hazra}, Gopal and {Jiang}, Jie and {Karak}, Bidya Binay and {Kitchatinov}, Leonid},
        title = "{Exploring the Cycle Period and Parity of Stellar Magnetic Activity with Dynamo Modeling}",
      journal = {\apj},
         year = 2019,
        month = oct,
       volume = {884},
       number = {1},
          eid = {35},
        pages = {35},
          doi = {10.3847/1538-4357/ab4128},
archivePrefix = {arXiv},
       eprint = {1909.01286},
 primaryClass = {astro-ph.SR},
       adsurl = {https://ui.adsabs.harvard.edu/abs/2019ApJ...884...35H}
}

@ARTICLE{Kumar22,
       author = {{Kumar}, Pawan and {Biswas}, Akash and {Karak}, Bidya Binay},
        title = "{Physical link of the polar field buildup with the Waldmeier effect broadens the scope of early solar cycle prediction: Cycle 25 is likely to be slightly stronger than Cycle 24}",
      journal = {\mnras},
         year = 2022,
        month = jun,
       volume = {513},
       number = {1},
        pages = {L112-L116},
          doi = {10.1093/mnrasl/slac043},
archivePrefix = {arXiv},
       eprint = {2203.11494},
 primaryClass = {astro-ph.SR},
       adsurl = {https://ui.adsabs.harvard.edu/abs/2022MNRAS.513L.112K}
}

@ARTICLE{KR99,
   author = {{Kitchatinov}, L.~L. and {R{\"u}diger}, G.},
    title = "{Differential rotation models for late-type dwarfs and giants}",
  journal = {\aap},
     year = 1999,
    month = apr,
   volume = 344,
    pages = {911-917},
   adsurl = {https://ui.adsabs.harvard.edu/abs/1999A%26A...344..911K}
}

@ARTICLE{KMB18,
       author = {{Karak}, Bidya Binay and {Mandal}, Sudip and {Banerjee}, Dipankar},
        title = "{Double Peaks of the Solar Cycle: An Explanation from a Dynamo Model}",
      journal = {\apj},
         year = 2018,
        month = oct,
       volume = {866},
       number = {1},
          eid = {17},
        pages = {17},
          doi = {10.3847/1538-4357/aada0d},
archivePrefix = {arXiv},
       eprint = {1808.03922},
 primaryClass = {astro-ph.SR},
       adsurl = {https://ui.adsabs.harvard.edu/abs/2018ApJ...866...17K}
}

@ARTICLE{kumar21b,
       author = {{Kumar}, Pawan and {Karak}, Bidya Binay and {Vashishth}, Vindya},
        title = "{Supercriticality of the Dynamo Limits the Memory of the Polar Field to One Cycle}",
      journal = {\apj},
         year = 2021,
        month = may,
       volume = {913},
       number = {1},
          eid = {65},
        pages = {65},
          doi = {10.3847/1538-4357/abf0a1},
archivePrefix = {arXiv},
       eprint = {2103.11754},
 primaryClass = {astro-ph.SR},
       adsurl = {https://ui.adsabs.harvard.edu/abs/2021ApJ...913...65K}
}

@ARTICLE{Kar23,
       author = {{Karak}, Bidya Binay},
        title = "{Models for the long-term variations of solar activity}",
      journal = {Living Reviews in Solar Physics},
         year = 2023,
        month = dec,
       volume = {20},
       number = {1},
          eid = {3},
        pages = {3},
          doi = {10.1007/s41116-023-00037-y},
archivePrefix = {arXiv},
       eprint = {2305.17188},
 primaryClass = {astro-ph.SR},
       adsurl = {https://ui.adsabs.harvard.edu/abs/2023LRSP...20....3K}
}

@ARTICLE{Kap16,
   author = {{K{\"a}pyl{\"a}}, M.~J. and {K{\"a}pyl{\"a}}, P.~J. and {Olspert}, N. and 
	{Brandenburg}, A. and {Warnecke}, J. and {Karak}, B.~B. and 
	{Pelt}, J.},
    title = "{Multiple dynamo modes as a mechanism for long-term solar activity variations}",
  journal = {\aap},
archivePrefix = "arXiv",
   eprint = {1507.05417},
 primaryClass = "astro-ph.SR",
     year = 2016,
    month = may,
   volume = 589,
      eid = {A56},
    pages = {A56},
      doi = {10.1051/0004-6361/201527002},
   adsurl = {http://adsabs.harvard.edu/abs/2016A%26A...589A..56K}
}

@ARTICLE{KC16,
   author = {{Karak}, B.~B. and {Cameron}, R.},
    title = "{Babcock-Leighton Solar Dynamo: The Role of Downward Pumping and the Equatorward Propagation of Activity}",
  journal = {\apj},
archivePrefix = "arXiv",
   eprint = {1605.06224},
 primaryClass = "astro-ph.SR",
     year = 2016,
    month = nov,
   volume = 832,
      eid = {94},
    pages = {94},
      doi = {10.3847/0004-637X/832/1/94},
   adsurl = {http://adsabs.harvard.edu/abs/2016ApJ...832...94K}
}

@ARTICLE{KM17,
   author = {{Karak}, B.~B. and {Miesch}, M.},
    title = "{Solar Cycle Variability Induced by Tilt Angle Scatter in a Babcock-Leighton Solar Dynamo Model}",
  journal = {\apj},
archivePrefix = "arXiv",
   eprint = {1706.08933},
 primaryClass = "astro-ph.SR",
     year = 2017,
    month = sep,
   volume = 847,
      eid = {69},
    pages = {69},
      doi = {10.3847/1538-4357/aa8636},
   adsurl = {http://adsabs.harvard.edu/abs/2017ApJ...847...69K}
}

@ARTICLE{KC11,
   author = {{Karak}, B.~B. and {Choudhuri}, A.~R.},
    title = "{The Waldmeier effect and the flux transport solar dynamo}",
  journal = {\mnras},
archivePrefix = "arXiv",
   eprint = {1008.0824},
 primaryClass = "astro-ph.SR",
     year = 2011,
    month = jan,
   volume = 410,
    pages = {1503-1512},
      doi = {10.1111/j.1365-2966.2010.17531.x},
   adsurl = {http://adsabs.harvard.edu/abs/2011MNRAS.410.1503K}
}

@ARTICLE{KKC14,
   author = {{Karak}, B.~B. and {Kitchatinov}, L.~L. and {Choudhuri}, A.~R.
	},
    title = "{A Dynamo Model of Magnetic Activity in Solar-like Stars with Different Rotational Velocities}",
  journal = {\apj},
archivePrefix = "arXiv",
   eprint = {1402.1874},
 primaryClass = "astro-ph.SR",
     year = 2014,
    month = aug,
   volume = 791,
      eid = {59},
    pages = {59},
      doi = {10.1088/0004-637X/791/1/59},
   adsurl = {http://adsabs.harvard.edu/abs/2014ApJ...791...59K}
}

@ARTICLE{KO11b,
   author = {{Kitchatinov}, L.~L. and {Olemskoy}, S.~V.},
    title = "{Differential rotation of main-sequence dwarfs and its dynamo efficiency}",
  journal = {\mnras},
archivePrefix = "arXiv",
   eprint = {1009.3734},
 primaryClass = "astro-ph.SR",
     year = 2011,
    month = feb,
   volume = 411,
    pages = {1059-1066},
      doi = {10.1111/j.1365-2966.2010.17737.x},
   adsurl = {http://adsabs.harvard.edu/abs/2011MNRAS.411.1059K}
}

@ARTICLE{KO12b,
       author = {{Kitchatinov}, L.~L. and {Olemskoy}, S.~V.},
        title = "{Differential rotation of main-sequence dwarfs: predicting the dependence on surface temperature and rotation rate}",
      journal = {\mnras},
         year = 2012,
        month = jul,
       volume = {423},
       number = {4},
        pages = {3344-3351},
          doi = {10.1111/j.1365-2966.2012.21126.x},
archivePrefix = {arXiv},
       eprint = {1204.4261},
 primaryClass = {astro-ph.SR},
       adsurl = {https://ui.adsabs.harvard.edu/abs/2012MNRAS.423.3344K}
}

@BOOK{KR80,
    author = {{Krause}, F. and {R{\"a}dler}, K.~H.},
    title = "{Mean-field magneto\-hydro\-dynamics and dynamo theory}",
    publisher = {Oxford: Pergamon Press},
    year = 1980}

@ARTICLE{LB89,
   author = {{Lean}, J.~L. and {Brueckner}, G.~E.},
    title = "{Intermediate-term solar periodicities - 100-500 days}",
  journal = {\apj},
     year = 1989,
    month = feb,
   volume = 337,
    pages = {568-578},
      doi = {10.1086/167124},
   adsurl = {http://adsabs.harvard.edu/abs/1989ApJ...337..568L}
}

@ARTICLE{NM07,
       author = {{Nandy}, Dibyendu and {Martens}, P.~C.~H.},
        title = "{Space Climate and the Solar Stellar connection: What can we learn from the stars about long-term solar variability?}",
      journal = {Advances in Space Research},
         year = 2007,
        month = jan,
       volume = {40},
       number = {7},
        pages = {891-898},
          doi = {10.1016/j.asr.2007.01.079},
       adsurl = {https://ui.adsabs.harvard.edu/abs/2007AdSpR..40..891N}
}

@ARTICLE{Noyes84a,
   author = {{Noyes}, R.~W. and {Hartmann}, L.~W. and {Baliunas}, S.~L. and 
	{Duncan}, D.~K. and {Vaughan}, A.~H.},
    title = "{Rotation, convection, and magnetic activity in lower main-sequence stars}",
  journal = {\apj},
     year = 1984,
    month = apr,
   volume = 279,
    pages = {763-777},
      doi = {10.1086/161945},
   adsurl = {http://adsabs.harvard.edu/abs/1984ApJ...279..763N}
}

@ARTICLE{Ol16,
       author = {{Ol{\'a}h}, K. and {K{\H{o}}v{\'a}ri}, Zs. and {Petrovay}, K. and
         {Soon}, W. and {Baliunas}, S. and {Koll{\'a}th}, Z. and {Vida}, K.},
        title = "{Magnetic cycles at different ages of stars}",
      journal = {\aap},
         year = "2016",
        month = "Jun",
       volume = {590},
          eid = {A133},
        pages = {A133},
          doi = {10.1051/0004-6361/201628479},
archivePrefix = {arXiv},
       eprint = {1604.06701},
 primaryClass = {astro-ph.SR},
       adsurl = {https://ui.adsabs.harvard.edu/abs/2016A&A...590A.133O}
}

@ARTICLE{Paxton,
       author = {{Paxton}, Bill},
        title = "{EZ to Evolve ZAMS Stars: A Program Derived from Eggleton's Stellar Evolution Code}",
      journal = {\pasp},
         year = 2004,
        month = jul,
       volume = {116},
       number = {821},
        pages = {699-701},
          doi = {10.1086/422345},
archivePrefix = {arXiv},
       eprint = {astro-ph/0405130},
 primaryClass = {astro-ph},
       adsurl = {https://ui.adsabs.harvard.edu/abs/2004PASP..116..699P}
}

@ARTICLE{Petrovay20,
       author = {{Petrovay}, Krist{\'o}f},
        title = "{Solar cycle prediction}",
      journal = {Living Reviews in Solar Physics},
         year = 2020,
        month = mar,
       volume = {17},
       number = {1},
          eid = {2},
        pages = {2},
          doi = {10.1007/s41116-020-0022-z},
archivePrefix = {arXiv},
       eprint = {1907.02107},
 primaryClass = {astro-ph.SR},
       adsurl = {https://ui.adsabs.harvard.edu/abs/2020LRSP...17....2P}
}

@ARTICLE{R84,
    author = {{Rengarajan}, T.~N.},
    title = "{Age-rotation relationship for late-type main-sequence stars}",
    journal = {\apjl},
    year = 1984,
    volume = 283,
    pages = {L63-L65}}

@ARTICLE{Shah18,
       author = {{Shah}, Shivani P. and {Wright}, Jason T. and {Isaacson}, Howard and {Howard}, Andrew W. and {Curtis}, Jason L.},
        title = "{HD 4915: A Maunder Minimum Candidate}",
      journal = {\apjl},
         year = 2018,
        month = aug,
       volume = {863},
       number = {2},
          eid = {L26},
        pages = {L26},
          doi = {10.3847/2041-8213/aad40c},
archivePrefix = {arXiv},
       eprint = {1801.09650},
 primaryClass = {astro-ph.SR},
       adsurl = {https://ui.adsabs.harvard.edu/abs/2018ApJ...863L..26S}
}

@ARTICLE{Skumanich72,
       author = {{Skumanich}, A.},
        title = "{Time Scales for CA II Emission Decay, Rotational Braking, and Lithium Depletion}",
      journal = {\apj},
         year = 1972,
        month = feb,
       volume = {171},
        pages = {565},
          doi = {10.1086/151310},
       adsurl = {https://ui.adsabs.harvard.edu/abs/1972ApJ...171..565S}
}

@ARTICLE{SB02,
       author = {{Saar}, S.~H. and {Brandenburg}, A.},
        title = "{A new look at dynamo cycle amplitudes}",
      journal = {Astronomische Nachrichten},
         year = 2002,
        month = jul,
       volume = {323},
        pages = {357-360},
          doi = {10.1002/1521-3994(200208)323:3/4<357::AID-ASNA357>3.0.CO;2-I},
archivePrefix = {arXiv},
       eprint = {astro-ph/0207392},
 primaryClass = {astro-ph},
       adsurl = {https://ui.adsabs.harvard.edu/abs/2002AN....323..357S}
}

@ARTICLE{Sch13,
       author = {{Schr{\"o}der}, K. -P. and {Mittag}, M. and {Hempelmann}, A. and
         {Gonz{\'a}lez-P{\'e}rez}, J.~N. and {Schmitt}, J.~H.~M.~M.},
        title = "{What do the Mt. Wilson stars tell us about solar activity?}",
      journal = {\aap},
         year = "2013",
        month = "Jun",
       volume = {554},
          eid = {A50},
        pages = {A50},
          doi = {10.1051/0004-6361/201219830},
       adsurl = {https://ui.adsabs.harvard.edu/abs/2013A&A...554A..50S}
}

@ARTICLE{SBZ94,
   author = {{Soon}, W.~H. and {Baliunas}, S.~L. and {Zhang}, Q.},
    title = "{Variations in surface activity of the Sun and solar-type stars}",
  journal = {\solphys},
     year = 1994,
    month = oct,
   volume = 154,
    pages = {385-391},
      doi = {10.1007/BF00681107},
   adsurl = {https://ui.adsabs.harvard.edu/abs/1994SoPh..154..385S}
}

@ARTICLE{SBZ93,
   author = {{Soon}, W.~H. and {Baliunas}, S.~L. and {Zhang}, Q.},
    title = "{An interpretation of cycle periods of stellar chromospheric activity}",
  journal = {\apjl},
     year = 1993,
    month = sep,
   volume = 414,
    pages = {L33-L36},
      doi = {10.1086/186989},
   adsurl = {https://ui.adsabs.harvard.edu/abs/1993ApJ...414L..33S}
}

@ARTICLE{TC23,
       author = {{Thibeault}, Christian and {Miara}, Lo{\"\i}c and {Charbonneau}, Paul},
        title = "{Nonlinearity, time delay, and Grand Maxima in supercritical Babcock-Leighton dynamos}",
      journal = {Journal of Space Weather and Space Climate},
         year = 2023,
        month = nov,
       volume = {13},
          eid = {32},
        pages = {32},
          doi = {10.1051/swsc/2023029},
       adsurl = {https://ui.adsabs.harvard.edu/abs/2023JSWSC..13...32T}
}

@ARTICLE{Viodoto14,
       author = {{Vidotto}, A.~A. and {Gregory}, S.~G. and {Jardine}, M. and {Donati}, J.~F. and {Petit}, P. and {Morin}, J. and {Folsom}, C.~P. and {Bouvier}, J. and {Cameron}, A.~C. and {Hussain}, G. and {Marsden}, S. and {Waite}, I.~A. and {Fares}, R. and {Jeffers}, S. and {do Nascimento}, J.~D.},
        title = "{Stellar magnetism: empirical trends with age and rotation}",
      journal = {\mnras},
         year = 2014,
        month = jul,
       volume = {441},
       number = {3},
        pages = {2361-2374},
          doi = {10.1093/mnras/stu728},
archivePrefix = {arXiv},
       eprint = {1404.2733},
 primaryClass = {astro-ph.SR},
       adsurl = {https://ui.adsabs.harvard.edu/abs/2014MNRAS.441.2361V}
}

@ARTICLE{VKK23,
       author = {{Vashishth}, Vindya and {Karak}, Bidya Binay and {Kitchatinov}, Leonid},
        title = "{Dynamo modelling for cycle variability and occurrence of grand minima in Sun-like stars: rotation rate dependence}",
      journal = {\mnras},
         year = 2023,
        month = jun,
       volume = {522},
       number = {2},
        pages = {2601-2610},
          doi = {10.1093/mnras/stad1105},
archivePrefix = {arXiv},
       eprint = {2304.05819},
 primaryClass = {astro-ph.SR},
       adsurl = {https://ui.adsabs.harvard.edu/abs/2023MNRAS.522.2601V}
}

@ARTICLE{BV07,
   author = {{B{\"o}hm-Vitense}, E.},
    title = "{Chromospheric Activity in G and K Main-Sequence Stars, and What It Tells Us about Stellar Dynamos}",
  journal = {\apj},
     year = 2007,
    month = mar,
   volume = 657,
    pages = {486-493},
      doi = {10.1086/510482},
   adsurl = {http://adsabs.harvard.edu/abs/2007ApJ...657..486B}
}

@ARTICLE{wright11,
   author = {{Wright}, N.~J. and {Drake}, J.~J. and {Mamajek}, E.~E. and
        {Henry}, G.~W.},
    title = "{The Stellar-activity-Rotation Relationship and the Evolution of Stellar Dynamos}",
  journal = {\apj},
archivePrefix = "arXiv",
   eprint = {1109.4634},
 primaryClass = "astro-ph.SR",
     year = 2011,
    month = dec,
   volume = 743,
      eid = {48},
    pages = {48},
      doi = {10.1088/0004-637X/743/1/48},
   adsurl = {http://adsabs.harvard.edu/abs/2011ApJ...743...48W}
}

@ARTICLE{wald,
   author = {{Waldmeier}, M.},
    title = "{Neue Eigenschaften der Sonnenfleckenkurve}",
  journal = {Astronomische Mitteilungen der Eidgen{\"o}ssischen Sternwarte Zurich},
     year = 1935,
   volume = 14,
    pages = {105-136},
   adsurl = {http://adsabs.harvard.edu/abs/1935MiZur..14..105W}
}

@phdthesis{Egeland:2017:thesis,
  Adsurl = {http://adsabs.harvard.edu/abs/2017PhDT.........3E},
  Author = {{Egeland}, R.},
  Month ={April},
  School = {Montana State University, Bozeman, Montana, USA},
  Title = {{Long-Term Variability of the Sun in the Context of Solar-Analog Stars}},
  Url = {https://scholarworks.montana.edu/xmlui/handle/1/12774},
  Year = 2017
}


\end{document}